\documentclass[prc,aps,twocolumn,showpacs,amssymb,superscriptaddress,fleqn]{revtex4-1}

\usepackage{color}
\usepackage[table]{xcolor}
\usepackage[english]{babel}
\usepackage{graphicx}
\usepackage{dcolumn}
\usepackage{mathtools}
\usepackage{relsize}
\usepackage{float}
\usepackage{braket}
\usepackage[margin=0.7in,columnsep=0.28in]{geometry}
\usepackage{amsmath,amssymb,bm}
\usepackage{microtype}
\usepackage{xcolor}
\usepackage{textcomp}
\usepackage[most]{tcolorbox}
\usepackage{booktabs}
\usepackage[colorlinks=true,citecolor=blue,linkcolor=blue]{hyperref}

\newcommand{\thetaeff}{$\Theta_{\rm eff}$}
\newcommand{\thetaeffs}{$\Theta_{\rm eff}$'s}

\newcommand{\zbb}{$0\nu\beta\beta$}

\newcommand{\heff}{$H_{\rm eff}$}
\newcommand{\heffs}{$H_{\rm eff}$'s}
\newcommand{\qbox}{$\hat{Q}$~box}
\newcommand{\tbox}{$\hat{\Theta}$~box}
\newcommand{\nme}{$M^{0\nu}$}
\newcommand{\nmel}{$M^{0\nu}_{\rm L}$}
\newcommand{\nmess}{$M^{0\nu}_{\rm S}$}
\newcommand{\nmes}{$M^{0\nu}$'s}

\newcommand{\nmeds}{$M^{2\nu}$'s}

\renewcommand{\vec}[1]{\mathbf{#1}}
\newcommand{\svec}[1]{\boldsymbol{#1}}
\newcommand{\isovec}[1]{\bm{#1}}

\newcommand{\dd}{\mathrm{d}}

\newcommand{\la}{\langle}
\newcommand{\ra}{\rangle}
\newcommand{\vsigma}{\svec{\sigma}}
\newcommand{\vtau}{\isovec{\tau}}

\newcommand{\NNNLO}{N$^3$LO}
\newcommand{\Bino}[2]{\ensuremath{C^{#2}_{#1}}}

\newcommand{\CG}[6]{\ensuremath{C^{#5#6}_{#1#2,#3#4}}}

\newcommand{\SixJ}[6]{\left\{\!\!%
\begin{array}{ccc}%
{#1} & {#2} & {#3}\\[0.2em]%
{#4} & {#5} & {#6}%
\end{array}%
\!\!\right\}%
}

\newcommand{\NineJ}[9]{\left\{\!\!%
\begin{array}{ccc}%
{#1} & {#2} & {#3}\\[0.2em]%
{#4} & {#5} & {#6}\\[0.2em]%
{#7} & {#8} & {#9}%
\end{array}%
\!\!\right\}%
}
\newcommand{\permille}{\text{\textperthousand}}
\begin{document}

\title{The renormalization of the shell-model neutrinoless
  double-$\beta$ decay operator starting from effective field theory
  at next-to-next-to-leading order}

\author{S. L. Lyu}
\affiliation{Dipartimento di Matematica e Fisica, Universit\`a degli
  Studi della Campania ``Luigi Vanvitelli'', viale Abramo Lincoln 5 -
  I-81100 Caserta, Italy}
\affiliation{Istituto Nazionale di Fisica Nucleare, \\ 
Complesso Universitario di Monte  S. Angelo, Via Cintia - I-80126
Napoli, Italy}
\affiliation{Department of Physics, Hunan Normal University, Changsha
  410081, China}
\author{U. Carusone}
\affiliation{Dipartimento di Matematica e Fisica, Universit\`a degli
  Studi della Campania ``Luigi Vanvitelli'', viale Abramo Lincoln 5 -
  I-81100 Caserta, Italy}
\author{G. De Gregorio}
\affiliation{Dipartimento di Matematica e Fisica, Universit\`a degli
  Studi della Campania ``Luigi Vanvitelli'', viale Abramo Lincoln 5 -
  I-81100 Caserta, Italy}
\affiliation{Istituto Nazionale di Fisica Nucleare, \\ 
Complesso Universitario di Monte  S. Angelo, Via Cintia - I-80126 Napoli, Italy}
\author{N. Itaco}
\affiliation{Dipartimento di Matematica e Fisica, Universit\`a degli
  Studi della Campania ``Luigi Vanvitelli'', viale Abramo Lincoln 5 -
  I-81100 Caserta, Italy}
\affiliation{Istituto Nazionale di Fisica Nucleare, \\ 
Complesso Universitario di Monte  S. Angelo, Via Cintia - I-80126 Napoli, Italy}
\author{L. Coraggio}
\affiliation{Dipartimento di Matematica e Fisica, Universit\`a degli
  Studi della Campania ``Luigi Vanvitelli'', viale Abramo Lincoln 5 -
  I-81100 Caserta, Italy}
\affiliation{Istituto Nazionale di Fisica Nucleare, \\ 
Complesso Universitario di Monte  S. Angelo, Via Cintia - I-80126 Napoli, Italy}

\begin{abstract}
This paper is the continuation of a previous study, where shell-model
calculations of the matrix elements for the neutrinoless double-$\beta$
decay were performed within a fully-consistent framework, starting
from expressions of the nuclear Hamiltonian and decay operators which
have been derived through chiral perturbation theory.
The decays under our investigation are those involving $^{48}$Ca,
$^{76}$Ge, and $^{82}$Se, as in our earlier analyses of electroweak
nuclear decays with Hamiltonians and operators derived within effective
field theory.
In this work we focus on perturbative properties of the chiral
expansion of the decay operator, namely in assessing the weight of
terms beyond the leading order on the nuclear matrix elements of
medium-mass candidates.
To this end, we have constructed effective shell-model decay operators
starting from two-body currents up to next-to-next-to-leading
order and, for the first time, we introduce for nuclei of experimental
interest contributions of the three-body current operators, which
emerge at next-to-next-to-next-to-leading order of the chiral
perturbation theory.
Our investigation is aimed to evaluate the theoretical uncertainties
affecting the calculated nuclear matrix elements of the neutrinoless
double-$\beta$ decay, by conjugating the estimate of the inaccuracy
related to the finiteness of the degrees of freedom of the nuclear
model and the truncation of the chiral expansion of the decay
operator.
\end{abstract}


\maketitle
\section{Introduction}\label{intro}
The introduction of the effective field theory (EFT) to solve the nuclear
many-body problem represents a kind of Copernican revolution, since
the task to build nuclear Hamiltonians and transition operators, as well
as the identification of the necessary degrees of freedom, has shifted
from the phenomenology of nuclear systems to the underlying theory of
the strong force, namely the quantum chromodynamics (QCD)
\cite{Weinberg79,Weinberg90,Weinberg91}.
This has been carried out by taking advantage of the separation
of energy scales characterizing the particles which are subject to the
strong force, namely quark and gluons, hadrons, and finally atomic
nuclei.

This new perspective to study nuclear physics has also involved the
calculation of nuclear matrix elements of electroweak processes, and,
in particular, those relating the wave functions of parent and
grand-daughter nuclei in the neutrinoless double-$\beta$ (\zbb) decay.

Recently, major efforts have been devoted to calculate terms of the
chiral perturbative expansion of the currents which structure the
\zbb-decay operator within a defined set of degrees of freedom --
namely nucleons and pions --, and a chosen energy scale.
At present, the expansion of the \zbb~ two-body current has been
performed up to next-to-next-to-leading order (N$^2$LO)
\cite{Cirigliano18a,Cirigliano19} in the chiral perturbation theory
(ChPT), and more recently also the lowest-order terms of the
three-body currents have been derived, which in a $\Delta$-less
approach start to contribute at next-to-next-to-next-to-leading order
(N$^3$LO) \cite{Chambers-Wall26}.

The aim of our research work is to study the issues arising from a
shell-model (SM) calculation of the nuclear matrix elements of the
\zbb~ decay (\nme), employing a nuclear Hamiltonian and transition
operators derived consistently within the same ChPT framework.
Starting from such premises, we consider a two-nucleon force (2NF)
derived through ChPT including contributions up to N$^3$LO
\cite{Entem03}, paired with a three-nucleon component that has been
consistently built at N$^2$LO \cite{Machleidt11}.
Then, effective SM Hamiltonians (\heffs) and decay operators
(\thetaeffs) are built by way of many-body perturbation theory to
account for the nuclear configurations of the full many-body
Hamiltonian, which are not explicitly included in the truncated model
space where only the valence nucleons interact outside the closed
core \cite{Kuo81,Suzuki95,Coraggio12a,Coraggio20c,Coraggio24b}.
Within our framework, these are the input elements for the SM
calculations of the \nmes~ for the decay of $^{48}$Ca, $^{76}$Ge, and
$^{82}$Se.

In a previous work, we have considered an expression of the decay
operator -- within the light Majorana-neutrino exchange --, that
includes only the LO contribution of the ChPT expansion to the
\zbb-decay two-body currents \cite{Coraggio26}.
This LO term is characterized by a long-range component, that is
substantially equal to one employed in the standard formulation of the
operator \cite{Tomoda91,Vergados12}, but also by a short-range one,
that consists of a contact operator.
This term is required to guarantee the renormalizability of the process
\cite{Cirigliano18a,Cirigliano19}, and introduces a new low-energy
constant (LEC) that has to be renormalized by evaluating the $nn
\rightarrow pp e^- e^-$ amplitude\cite{Cirigliano21}.

There, we have focussed our attention on the convergence properties of
both \heffs~ and \thetaeffs~ with respect to their many-body
perturbative expansion, mainly to provide an estimate of the
theoretical uncertainties affecting the calculated \nmes.

Here, we aim to evaluate the impact of higher-order terms of the ChPT
expansion of the \zbb-decay operator on the predicted \nmes~ of
$^{48}$Ca, $^{76}$Ge, and $^{82}$Se decays.
A similar investigation was carried out in Ref. \cite{Castillo25},
where the authors performed the calculation of \nmes~ within the nuclear
shell model and proton-neutron quasiparticle random-phase
approximation (pnQRPA), but employing phenomenological \heffs.

Since the \heffs~ were not derived within ChPT and there was not a
consistent link between them and the choice of the LECs which
characterize the two-body currents, the results of the \nmes~ were
presented varying the values of such LECs within a certain range.

This is at variance with our approach, because in the current study
the two-body currents LECs are the same as those defining the ChPT
nuclear Hamiltonian.

In addition to the two-body currents of the \zbb-decay operator, we
consider also the contributions of three-body currents, derived at
N$^3$LO of the ChPT expansion \cite{Chambers-Wall26}, and for the
first time they are included in the calculation of the \nmes~ for the
decay of nuclei of experimental interest.

Another issue that we are going to tackle is the choice of the LEC
characterizing the LO short-range contribution, usually dubbed as
$g_{\nu}^{\rm NN}$, that should be fixed by reproducing the data of
lepton-number-violation (LNV) processes, data that are not available
at present.
In Ref. \cite{Cirigliano19} the authors suggested that a possible
procedure to estimate the magnitude of $g_{\nu}^{\rm NN}$ is to relate
it to the electromagnetic charge-independence-breaking (CIB) contact
interactions, whose coefficients are constrained by nucleon-nucleon
scattering data. Although this relation does not provide a complete
determination of $g_{\nu}^{\rm NN}$, it can be used to estimate its natural
size.
This was the procedure that we employed in our previous work
\cite{Coraggio26}, and that can be found also in the paper by
Jokiniemi {\em et al} in Ref. \cite{Jokiniemi21}.

Here, we have additionally considered the value inferred by Wirth {\it et
  al} from calculations of the $nn \rightarrow pp e^-e^-$
transition amplitude performed with several chiral interactions,
including the Entem-Machleidt N$^3$LO 2NF \cite{Wirth21}.

As we will show, the two different approaches to the renormalization of
the LEC $g_{\nu}^{\rm NN}$, both defined for the 2NF considered to
construct the effective SM Hamiltonians, lead to very similar values
of the \nmes, a result that is quite relevant to assess the soundness
of the ChEFT approach to the definition of the \zbb-decay operator.

We have organized this work according to the following structure.
In Sec. \ref{theory}, we report some details about the expression
of the two-body currents of the \zbb~ decay operator up to N$^2$LO,
and of the three-body current at N$^3$LO.
A very short summary is also given of the procedure we have
followed to derive the effective SM Hamiltonian as well as the \zbb~
operator, starting from the chiral 2NF N$^3$LO potential constructed
by Entem and Machleidt \cite{Entem03} and supplemented by a three-body
component at N$^2$LO in ChPT \cite{Machleidt11}.
A more detailed description can be found in Ref. \cite{Coraggio26}.

The results of the SM calculations are then reported in
Sec. \ref{results}, where we present the outcome of the order-by-order
behavior of the \nmes~ with respect to the ChPT expansion of the
\zbb-decay operator.

Finally, the conclusions of our study are drawn in
Sec. \ref{conclusions}, as well as the perspectives of our future
projects.

\section{Theoretical framework}\label{theory}
As mentioned in the Introduction, we start -- for the present study --
from a nuclear Hamiltonian consisting of 2NF and 3NF components of the
nuclear force, both derived through a perturbative expansion of the
chiral Lagrangian at N$^3$LO and N$^2$LO, respectively
\cite{Entem03,Machleidt11}.

The 2NF and 3NF components share the same non-local regulator
function, with a cutoff $\Lambda=500$ MeV, as well as the LECs $c_1$,
$c_3$, and $c_4$, which appear in the two-pion exchange topology of
the 3NF \cite{Machleidt11}.
As regards the one-pion exchange and three-nucleon-contact
interactions of the 3NF, these terms are specified by the $c_D$ and
$c_E$ LECs, respectively, that are not present in the formulation of
the 2NF, and which should be fixed to reproduce the observables of the
$A=3$ system.
For these extra LECs we have adopted the same values as in
Refs. \cite{Fukui18,Ma19,Coraggio20e,Coraggio21,Coraggio24a,Lyu25a},
namely $c_D=-1$ and $c_E=-0.34$, which trace back to a study performed
in Ref. \cite{Navratil07a}.

Consistently with the approach we have followed for the nuclear
Hamiltonian, we have employed ChPT to construct also the electroweak
currents to construct the \zbb-decay operator for the light
Majorana-neutrino exchange.
EFT and ChPT allow to scale the physics of such a process from the
energy of the lepton-number violation $\Lambda_{\rm LNV}=1-100$ TeV
down to the chiral-symmetry breaking scale of low-energy nuclear
systems, namely the pion mass \cite{Cirigliano18a}.

The development of the chiral perturbative expansion of the \zbb-decay
operator has been recently carried out up to N$^2$LO for the two-body
currents \cite{Cirigliano18a,Cirigliano18b}, and also up to N$^3$LO
for the three-body ones \cite{Chambers-Wall26}.
Such an innovative approach has stimulated research studies in
evaluating \nmes~ for nuclei that are currently of experimental
interest \cite{Wirth21,Jokiniemi21,Belley24,Castillo25}, but the
impact of three-body currents has been assessed only for very light
nuclear systems of the $p$-shell region \cite{Chambers-Wall26}.

We have mentioned in the introduction that a relevant new feature that
characterizes the perturbative expansion of the \zbb~ decay operator
in ChPT is the emergence at LO of a short-range contact operator which is
not present in the usual formulation of the theory
\cite{Tomoda91,Vergados12}.
This term accomplishes the need to introduce a counterterm to absorb
the ultraviolet (UV) divergences that appear in the calculation of the
amplitude of the $nn \rightarrow pp e^-e^-$ process for $^1S_0
\rightarrow ^1S_0$ transitions \cite{Cirigliano19}.

Since this counterterm is regulated by a LEC, dubbed $g_{\nu}^{\rm
  NN}$, one has to face the question that the only way to fix it would
be by reproducing the data of LNV processes, data that are not
available at present.
Currently they have been suggested two related ways to estimate the
magnitude of the contact term, one was presented in
Ref. \cite{Cirigliano19}, where the authors suggested to consider
the coefficients of the charge-independence breaking contact
interaction.
Then, for the Entem-Machleidt N$^3$LO potential this leads to
$g_{\nu}^{\rm NN}=-0.47$ fm$^2$, if a local regulator is considered
for the contact term (see Ref. \cite{Jokiniemi21}), a value that we
have employed in our previous work \cite{Coraggio26}.

Here, we have also considered the approach followed by Wirth {\it et
  al}, who have calculated the $nn \rightarrow pp e^-e^-$
transition amplitude using different ChPT $NN$ interactions, and among
them they evaluated also the case for the Entem-Machleidt N$^3$LO 2NF
\cite{Wirth21}.
They have found that the synthetic datum can be reproduced with a
value of the LEC $g_{\nu}^{\rm NN}=-0.597$ fm$^2$, and a non-local
regulator, that is consistent with the one defined for the 2NF and 3NF
\cite{Entem03}.

Now, it is worth to sketch out some details of the structure of the
\zbb-decay operator for the light Majorana-neutrino exchange starting
from the expression of the half-life of such a process:

\begin{equation}
\left[ T^{0\nu}_{1/2}\right]^{-1} = G^{0\nu} g_A^4 \left| M^{0\nu} \right|^2
\left|\frac{ \langle m_{\nu}\rangle}{m_e}\right|^2~,
\label{halflife}
\end{equation}

\noindent
where $G^{0\nu}$ is the phase-space factor \cite{Kotila12,Kotila13},
\nme~ is the nuclear matrix element, $g_A$ is the axial  coupling
constant, $m_e$ is the electron mass, and $\langle m _{\nu} \rangle =
\sum_i (U_{ei})^2 m_i$ is the effective neutrino mass, as expressed in
terms of the neutrino masses $m_i$ and their mixing matrix elements
$U_{ei}$.

As we already showed in Ref. \cite{Coraggio26}, in the framework of
ChPT \nme~ is expressed as the sum of a long- and a short-range
component \cite{Cirigliano18a}:

\begin{equation}
M^{0\nu} = M^{0\nu}_{\rm L} + M^{0\nu}_{\rm S}
\label{nmeLS}
\end{equation}

We start from the LO short-range component \nmess~ that is expressed
in terms of the two-body transition-density matrix elements $\langle f |
a^{\dagger}_{p}a_{n} a^{\dagger}_{p^\prime} a_{n^\prime} | i \rangle$,
between the initial ($i$) and final ($f$) nuclear wave functions, as:

\begin{eqnarray}
M_{\rm S }^{0\nu}  & =  & \sum_{j_n j_{n^\prime} j_p j_{p^\prime}}
\langle f | a^{\dagger}_{p}a_{n} a^{\dagger}_{p^\prime} a_{n^\prime} 
| i \rangle \nonumber \\
~ & ~& \times  \left< j_p  j_{p^\prime} \mid \Theta_{\rm S} \mid  j_n j_{n^\prime}
       \right>~, \label{M0nuappCT}
\end{eqnarray}

\noindent
where $\Theta_{\rm S} $ is defined in terms of the neutrino potential
$H_{\rm S}$ as:

\begin{eqnarray}
\Theta_{\rm S} & = & \tau^-_{1} \tau^-_{2} H_{\rm S}(r)
                          \label{operatorCT} \, .
\end{eqnarray}
\noindent 

The expression of the neutrino potential $H_{\rm S}(r)$ employing a
local regulator can be found in our recent work \cite{Coraggio26},
while the one with a non-local regulator is reported in
Ref. \cite{Wirth21}.

As regards the expression of the long-range component \nmel~ of the
nuclear matrix element \nme, its expression has the same structure as
in Eq. (\ref{M0nuappCT}):

\begin{eqnarray}
M_{\rm L \alpha}^{0\nu} & =  & \sum_{j_n j_{n^\prime} j_p j_{p^\prime}}
\langle f | a^{\dagger}_{p}a_{n} a^{\dagger}_{p^\prime} a_{n^\prime} 
| i \rangle \nonumber \\
~ & ~& \times  \left< j_p  j_{p^\prime} \mid \Theta_{\rm L \alpha} \mid  j_n j_{n^\prime}
       \right>~, \label{M0nuappl}
\end{eqnarray}

\noindent
The label $\alpha$ stands for Fermi (F), Gamow-Teller (GT), or tensor
(T) decay channels, and the operators $\Theta_{\rm L \alpha}$ have the
following expression in terms of the neutrino potentials
$H_{\alpha}(r)$ \cite{Engel17}:

\begin{eqnarray}
 \Theta_{\rm GT} & = & [ \tau^-_{1} \tau^-_{2} (\vec{\sigma}_1 \cdot
                       \vec{\sigma}_2) H_{\rm GT}(r) ]\label{operatorGT} \, , \\
\Theta_{\rm F} & = & \tau^-_{1} \tau^-_{2} H_{\rm F}(r)
                          \label{operatorF} \, ,\\
\Theta_{\rm T} & = & [\tau^-_{1} \tau^-_{2} \left(
                         3\left(\vec{\sigma}_1 \cdot \hat{r} \right)
                         \left(\vec{\sigma}_2 \cdot \hat{r} \right) -
                         \right. \nonumber \\
  ~ & ~ & \left. \vec{\sigma}_1 \cdot \vec{\sigma}_2 \right) H_{\rm
          T}(r) ]~, \label{operatorT}
\end{eqnarray}

The neutrino potentials in turn are expressed in terms of the form
functions $h_{\alpha}$ as:

\begin{equation}
H_{\alpha}(r)=\frac {2R}{\pi} \int_{0}^{\infty} \frac {j_{n_{\alpha}}(qr)
  h_{\alpha} (q^2)qdq}{q}~,
\label{neutpotappl}
\end{equation}

\noindent
where the nuclear radius $R$ is defined as $R=1.2 A^{1/3}$ fm,
$j_{n_{\alpha}}(qr)$ is the spherical Bessel function, $n_{\alpha}=0$
for Fermi and Gamow-Teller components, and $n_{\alpha}=2$ for the
tensor component.

We refer to our previous work \cite{Coraggio26} for the formal
expression of the LO form functions, according to their
derivation in Ref. \cite{Cirigliano18a}.
Here, we provide some details on the structure of \nme at N$^2$LO of ChPT
-- in a $\Delta$-less theory there are no NLO contributions --, as well
as on the form functions $h_{\alpha}$.

We should point out that, up to N$^2$LO in the ChPT expansion, the
long-range component of nuclear matrix element  $M_{\rm L \alpha}^{0\nu}$ should be
decomposed for each channel $\alpha={\rm GT,F,T}$ as:
\begin{equation}
    M_{\rm L \alpha}^{0\nu}
    =
    M_{\rm{LO+1b}_{\alpha}}^{0\nu}
    +
    M_{\rm{loop}_{\alpha}}^{0\nu}
    +
    M_{\rm{usoft}_{\alpha}}^{0\nu},
\end{equation}
where $ M_{\rm{LO+1b}_{\alpha}}^{0\nu}$ denotes the long-range LO
contribution supplemented by the N$^2$LO corrections to the
single-nucleon currents, while $M_{\mathrm{loop}}^{0\nu}$ and 
$M_{\mathrm{usoft}}^{0\nu}$ denote the two-body loop and ultrasoft
contributions, respectively \cite{Cirigliano18a}.

The single-nucleon corrections preserve the Gamow-Teller, Fermi, and
tensor structure of the LO neutrino potential introduced in the
previous section, while accounting for the momentum dependence of the
nucleon weak form factors.
In particular, the corresponding Gamow-Teller, Fermi, and tensor form
functions become:

\begin{align}
h_{\rm GT}(q^2)
&=
g_A^2(q^2)
+
g_P(q^2)g_A(q^2)\frac{q^2}{3m_N}
\nonumber\\
&\quad
+
g_P^2(q^2)\frac{q^4}{12m_N^2}
+
g_M^2(q^2)\frac{q^2}{6g_A^2m_N^2},
\\
h_{\rm F}(q^2)
&=
g_V^2(q^2),
\\
h_{\rm T}(q^2)
&=
-g_P(q^2)g_A(q^2)\frac{q^2}{3m_N}
\nonumber\\
&\quad
-g_P^2(q^2)\frac{q^4}{12m_N^2}
+
g_M^2(q^2)\frac{q^2}{12g_A^2m_N^2}.
\end{align}

The vector and axial form factors are parametrized as

\begin{equation}
g_V(q^2)
=
\left(1+\frac{q^2}{\Lambda_V^2}\right)^{-2},
\qquad
g_A(q^2)
=
\left(1+\frac{q^2}{\Lambda_A^2}\right)^{-2},
\end{equation}

\noindent
with $\Lambda_V=850$ MeV and $\Lambda_A=1040$ MeV, while the magnetic
and induced-pseudoscalar form factors are given by

\begin{equation}
g_M(q^2)
=
(1+\kappa_1)g_V(q^2),
\qquad
g_P(q^2)
=
-\frac{2m_N g_A(q^2)}
{q^2+m_\pi^2},
\end{equation}

\noindent
where $\kappa_1=3.7$ is the nucleon isovector anomalous magnetic
moment.

The two-body contribution associated with a given component $X$ is

\begin{equation}
    M_{\mathrm{loop},X}^{0\nu}
    =
    \frac{4R}{\pi g_A^2}
    \left\langle 0_f^+
    \left\|
    \sum_{m,n}
    \tau_m^- \tau_n^-\,
    \mathcal{V}_X^{(m,n)}
    \right\|
    0_i^+
    \right\rangle ,
\end{equation}

\noindent
where $\mathcal{V}_X^{(m,n)}$ denotes the corresponding N$^2$LO
two-body neutrino potential, whose expression is reported in Appendix
\ref{app:neutrino-potentials}.
The label $X$ identifies the vector (VV), axial (AA), counterterm
(CT), and loop-ultrasoft term.

The expression of the ultrasoft contribution is

\begin{equation}
\begin{aligned}
M_{\mathrm{usoft}}^{0\nu}(\mu_{\mathrm{us}})
={}&
-\frac{2R}{\pi}
\sum_n
\langle 0_f^+|\bm{\sigma}\tau^-|n\rangle
\langle n|\bm{\sigma}\tau^-|0_i^+\rangle
\\[-1mm]
&\times
\left(
    \frac{Q_{\beta\beta}}{2}
    +m_e+E_n-E_i
\right)
\\[-1mm]
&\times
\left[
    \ln\left(
        \frac{\mu_{\mathrm{us}}}
        {2\left(
            Q_{\beta\beta}/2+m_e+E_n-E_i
        \right)}
    \right)
    +1
\right].
\end{aligned}
\label{eq:usoft}
\end{equation}

Here, $\mu_{\mathrm{us}}$ denotes the ultrasoft renormalization scale,
and $Q_{\beta\beta}$ is the $\beta\beta$-decay $Q$ value, whereas $E_i$
and $E_n$ denote the energies of the initial and intermediate nuclear
states, respectively.
As in Ref. \cite{Castillo25}, we assume that the two emitted electrons
share the available energy equally,

\begin{equation}
    E_{e_1}=E_{e_2}
    =
    \frac{Q_{\beta\beta}}{2}+m_e .
\end{equation}

\noindent
In Ref. \cite{Dekens24}, the authors showed that this approximation
introduces only a negligible error.

It is worth noting that three-body terms also arise at N$^2$LO when
the $\Delta$ isobar is included as an explicit degree of freedom
\cite{Chambers-Wall26}.
Since explicit $\Delta$-isobar excitations are not included in the
chosen nuclear Hamiltonian \cite{Entem03}, these three-body
contributions are not considered here.

As regards the contributions at N$^3$LO of the ChPT expansion of the
\zbb~ operator, we have considered only the contributions to the
three-body currents whose expressions have been derived in the
momentum-space representation in Ref. \cite{Chambers-Wall26}.
There, the authors derived also the terms of the ChPT expansion within
a $\Delta$-full approach, which start to appear already at N$^2$LO.

To employ \zbb~ three-body currents in a SM framework, the main step
is to calculate the three-body matrix elements of the three-nucleon
neutrino potentials between antisymmetrized harmonic-oscillator (HO)
basis states, consistently with the matrix elements of the nuclear
Hamiltonian and of the \zbb~ two-body currents.
The details of the procedure we have followed in present work are
reported in Appendix \ref{3BC}, and are based on the construction of
three-body matrix elements of the N$^2$LO 3NF we employ for the
construction of \heffs.

Now, we need to shift the attention to the derivation of the effective
SM Hamiltonians \heffs~ and \zbb-decay operators \thetaeffs~ which, as
in our previous studies was carried out within the framework of the
many-body perturbation theory \cite{Kuo71,Suzuki80,Suzuki95,Suzuki11}.

As reported in Ref. \cite{Coraggio20c}, both the derivation of \heff~
and \thetaeff~ are grounded on the perturbative expansion of vertex
functions, the \qbox~ \cite{Kuo71} and the \tbox~ \cite{Suzuki95},
respectively, a procedure that drives a consistent derivation of both
\heff~ and \thetaeff.

In our previous work in Ref. \cite{Coraggio26}, we have sketched out
the structure of these procedures and more details about the
perturbative expansion of \heffs~ and \thetaeffs~ can be found in
Refs. \cite{Coraggio12a,Coraggio20c}.
Here, we draw the attention on the treatment of the three-body
components of the effective SM \zbb-decay operator.

As a matter of fact, in the construction of \thetaeff~ we have to
consider two distinct sources of three-nucleon terms in its structure.

The first one are the aforementioned three-body currents which emerge
in the ChPT expansion of \zbb-decay operator.

The other comes from the fact that, even if a two-body structure only
is considered for the \zbb-decay operator, we are considering nuclear
systems with a number of valence nucleons far larger than two.
This means that for many valence-nucleons systems the perturbative
expansion of the \tbox~ includes also three-, four-, ..., many-body
diagrams which account for the progressive filling of model-space
orbitals.

This process is known as the so-called ``blocking effect'', urging to
take into account for the Pauli exclusion principle in systems with more
than two valence nucleons \cite{Towner87}.
We meet such an issue by including the leading terms of these
many-body contributions in the perturbative expansion of the \tbox~ of
second-order three-body diagrams, and details of such a procedure are
reported in Ref. \cite{Coraggio20c}.

Since the SM code we employ to calculate \nme~\cite{KSHELL} cannot
manage three-body operators, we treat the three-body components
of the \zbb-decay operator by deriving a density-dependent
two-body contribution at one-loop order from three-body diagrams,
summing over the partially-filled model-space orbitals.
The details of this procedure can be found in
Refs. \cite{Ma19,Coraggio20c}, where the same procedure was performed
to derive density-dependent two-body \heffs~ to diagonalize the SM
Hamiltonian for many-valence nucleon systems.

\section{Nuclear matrix elements for the $^{48}$Ca, $^{76}$Ge, and
  $^{82}$Se decays} \label{results}
Here, we discuss the results of the calculation of the \nmes~ for the
decay of $^{48}$Ca, $^{76}$Ge, and $^{82}$Se, employing two-body
currents derived up to N$^2$LO in ChPT and three-body currents at
N$^3$LO.

\begin{figure}[h]
\begin{center}
\includegraphics[scale=0.34,angle=0]{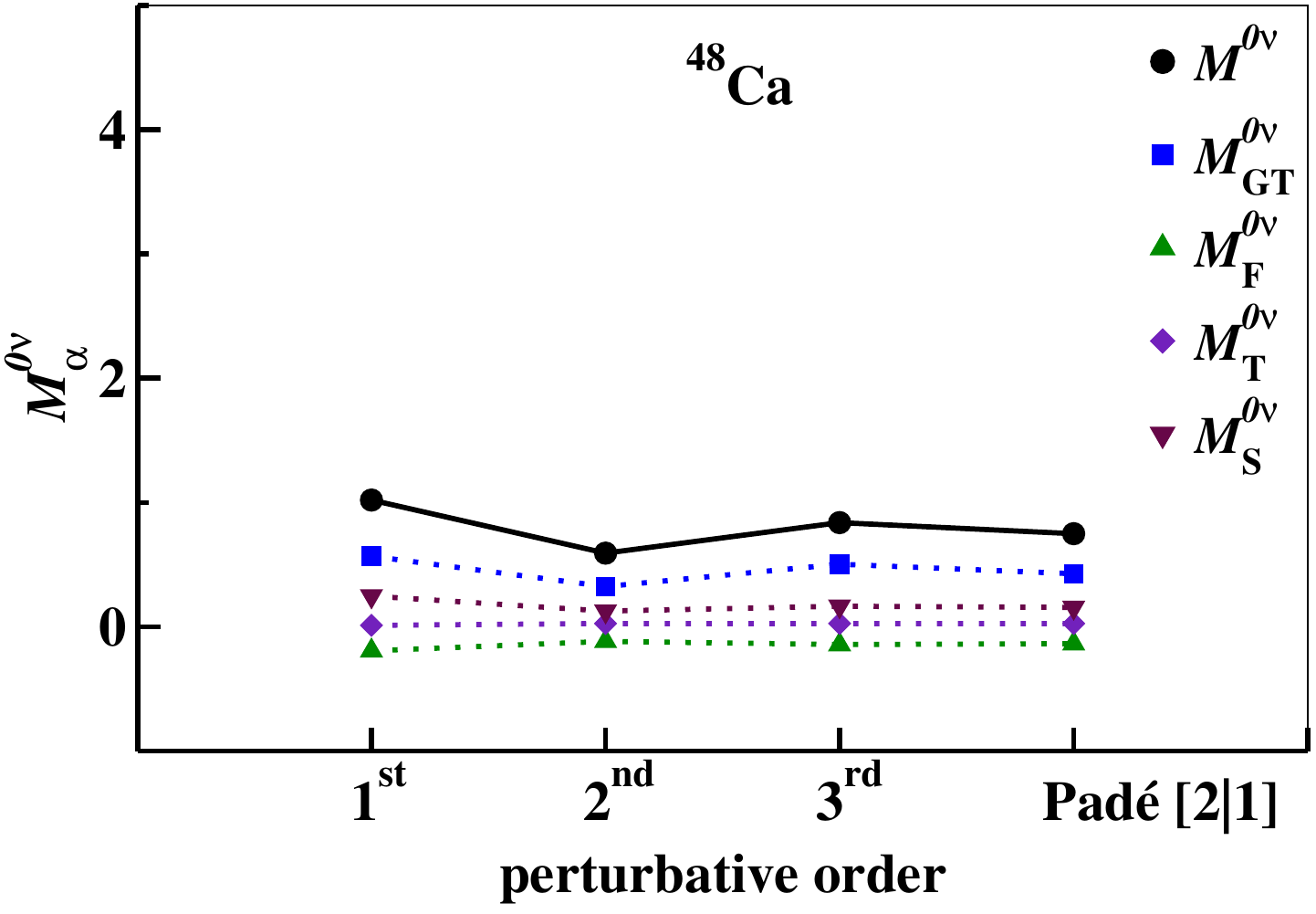}
\caption{\nme~ for the decay of the $^{48}$Ca ground state to the
  $^{48}$Ti one, as a function of the perturbative order. 
The green triangles correspond to $M^{0\nu}_{\rm F}$, the blue squares
to $M^{0\nu}_{\rm GT}$, the purple diamonds to $M^{0\nu}_{\rm T}$, the
brown lower triangles to $M^{0\nu}_{\rm S}$, and the black dots to
the full \nme.}
\label{48Ca_obo}
\end{center}
\end{figure}

First, we have to recall that in our previous works, where the same SM
Hamiltonians were used to calculate the nuclear wave functions, we
already reported a detailed study of the ability of reproducing the
observed spectroscopy of parent and grand-daughter nuclei that are
involved in such \zbb-decays, as well as the comparison between
experimental and calculated two-neutrino double-$\beta$ decay matrix
elements (\nmeds) \cite{Coraggio24a}.
In Ref. \cite{Coraggio26} it was also presented a study of the
perturbative behavior of the \heffs, comparing the low-lying
excitation spectra and electric-quadrupole strengths $B(E2)$s as
obtained with \heffs~ derived at first-, second-, and third-order in
many-body perturbation theory.

 These opening investigations were carried out to validate the quality of
our approach to calculate nuclear quantities by benchmarking theory vs
observables, showing the ability to reproduce quantitatively energy
spectra, the deviation of many-valence-nucleon systems from the SM
spherical shape, electroweak-process matrix elements, etc..
This is a mandatory step for any nuclear structure calculation to
support the predictive power of the calculated \nmes.

Now, it is worth starting from the discussion to assess the
perturbative behavior of our calculated \nmes~ with respect to the
many-body expansion of the effective SM operator \thetaeff, a step
that we made also in Ref. \cite{Coraggio26}.
This is necessary to evaluate the uncertainties affecting our \nmes~
because of the truncation of the \thetaeffs~ expansion at third order
of many-body perturbation theory.

\begin{figure}[t]
\begin{center}
\includegraphics[scale=0.34,angle=0]{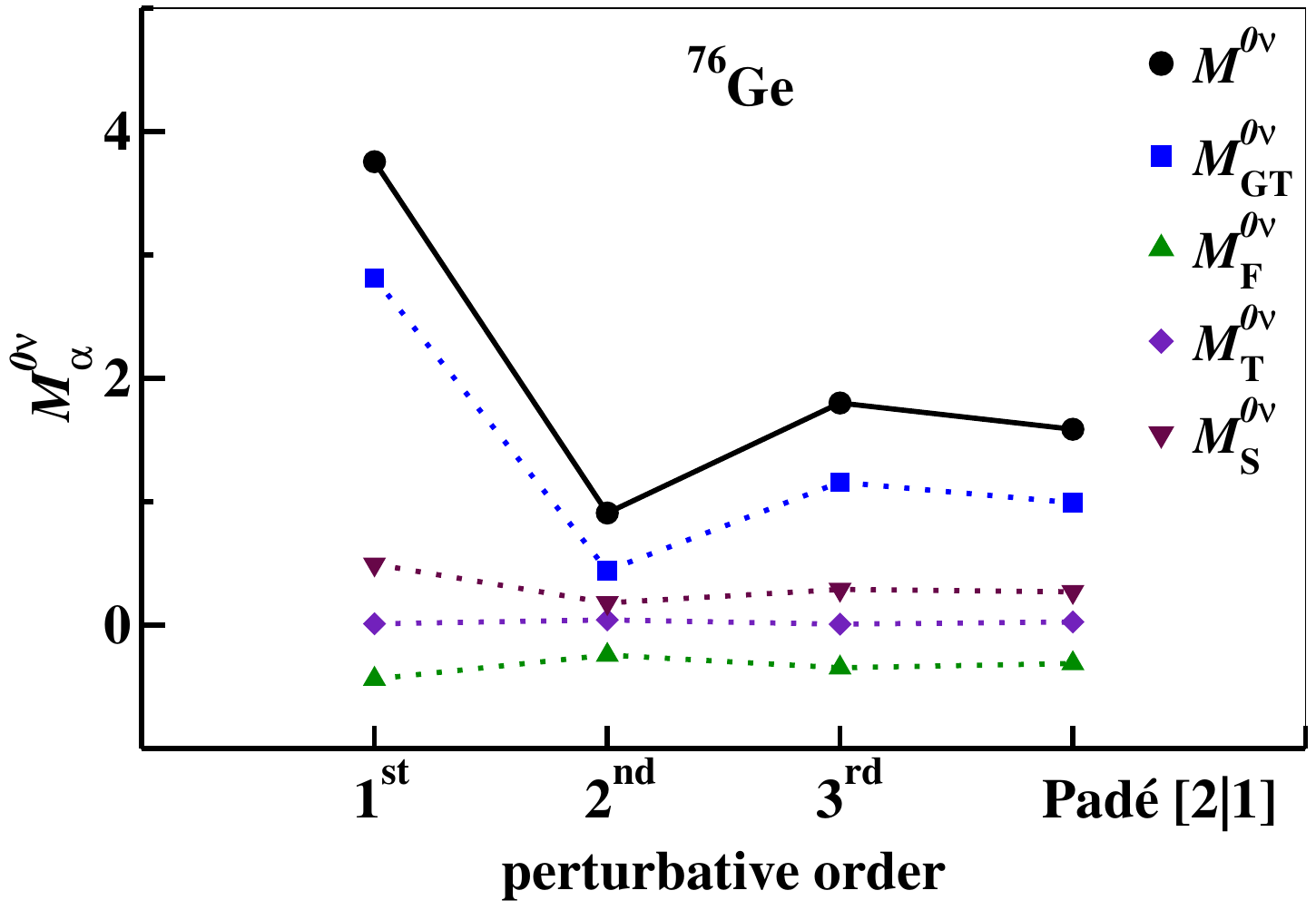}
\caption{Same as in Fig. \ref{48Ca_obo}, but for the decay of the
  $^{76}$Ge ground state to the $^{76}$Se one.}
\label{76Ge_obo}
\end{center}
\end{figure}

\begin{figure}[t]
\begin{center}
\includegraphics[scale=0.34,angle=0]{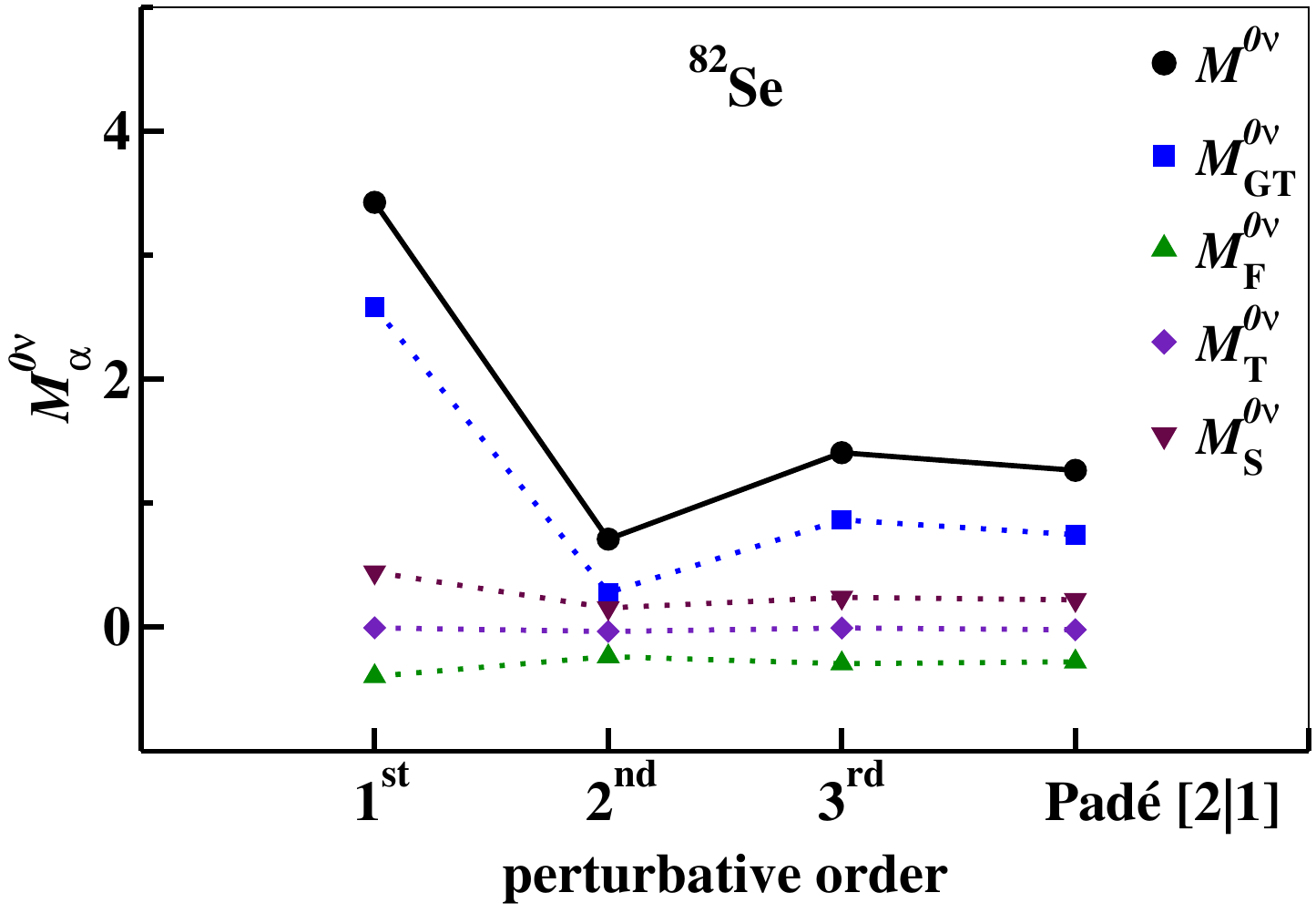}
\caption{Same as in Fig. \ref{48Ca_obo}, but for the decay of the
  $^{82}$Se ground state to the $^{82}$Kr one.}
\label{82Se_obo}
\end{center}
\end{figure}

We start considering the \zbb~ decay between the ground states of
$^{48}$Ca and $^{48}$Ti.
In Fig. \ref{48Ca_obo} we have reported the calculated values of the
\nmel~ components $M^{0\nu}_{\rm GT}$, $M^{0\nu}_{\rm F}$,
$M^{0\nu}_{\rm T}$, the short-range component \nmess~-- that has been
calculated using the same local regulator and $g_{\nu}^{\rm NN}$ value as in
Ref. \cite{Coraggio26} --, and the total \nme~ from first- up to
third-order in perturbation theory.
We report also the value of their Pad\'e approximant $[2|1]$, since it
is an indicator of the quality of the perturbative behavior
of the calculated \nme~ \cite{Baker70} and we will consider it as the
best estimate of our predicted \nmes.

\begin{table*}[t]
  \caption{Calculated values of \nme~ for all decays under
    investigation, employing two-body currents at LO and N$^2$LO in
    ChPT. The third and sixth columns correspond to the results obtained
    employing the effective SM \zbb-decay operator at third-order in
    perturbation theory, the fourth and seventh ones the corresponding
    Pad\'e approximant $[2|1]$ . In the fifth and eighth columns are
    the absolute value of the difference of results in the third and
    fourth columns, as well as the sixth and seventh ones,
    ($\hat{\Delta}$) are reported.}
\label{NME}
\setlength{\tabcolsep}{9pt}
\renewcommand{\arraystretch}{1.25}
\begin{tabular}{llccc|ccc}
\toprule
~ & ~ & \multicolumn{3}{c}{LO} & \multicolumn{3}{c}{N$^2$LO} \\
Decay & $J^{\pi}_i \rightarrow J^{\pi}_f$ &
$M^{0\nu}_{\rm 3rd}$ & $M^{0\nu}_{\textrm {Pad\'e}}$ & $\hat{\Delta}$ &
$M^{0\nu}_{\rm 3rd}$ & $M^{0\nu}_{\textrm {Pad\'e}}$ & $\hat{\Delta}$ \\
\midrule
  $^{48}$Ca  $\rightarrow$ $^{48}$Ti & ~ & ~ & ~ & ~ & ~ & ~ & ~ \\
~ & $0^+_1 \rightarrow 0^+_1$ & 0.89 & 0.80 & 0.09 & 0.84 & 0.75 & 0.09 \\
~ & $0^+_1 \rightarrow 0^+_2$ & 0.75 & 0.74 & 0.01 & 0.78 & 0.78 & 0.001 \\
\midrule
  $^{76}$Ge  $\rightarrow$ $^{76}$Se & ~ & ~ & ~ & ~ & ~ & ~ & ~ \\
 ~ & $0^+_1 \rightarrow 0^+_1$ & 1.66 & 1.46 & 0.20 & 1.80 & 1.59 & 0.21 \\
 ~ & $0^+_1 \rightarrow 0^+_2$ & 0.09 & 0.10 & 0.01 & 0.10 & 0.12 & 0.02 \\
\midrule
  $^{82}$Se  $\rightarrow$ $^{82}$Kr & ~ & ~ & ~ & ~ & ~ & ~ & ~ \\
 ~ & $0^+_1 \rightarrow 0^+_1$ & 1.25 & 1.12 & 0.13 & 1.41 & 1.26 & 0.15 \\
 ~ & $0^+_1 \rightarrow 0^+_2$ & 0.17 & 0.17 & 0.003 & 0.15 & 0.15 & 0.00001 \\
\bottomrule
\end{tabular}
\end{table*}

\begin{table*}[t]
  \caption{Calculated values of \nme~ for all decays under
    investigation employing both LO and N$^2$LO \zbb~ two-body
    currents. Between the parentheses they are reported the estimated
    uncertainty with respect to the perturbative expansion of the
    effective SM operator \thetaeff. The last column reports the
    percent difference between N$^2$LO and LO results.} 
\label{NMEobo}
\setlength{\tabcolsep}{9pt}
\renewcommand{\arraystretch}{1.25}
\begin{tabular}{llccc}
\toprule
Decay & $J^{\pi}_i \rightarrow J^{\pi}_f$ &  LO \nme & N$^2$LO \nme & Percent difference \\
\midrule
  $^{48}$Ca  $\rightarrow$ $^{48}$Ti & ~ & ~ & ~ & ~ \\
~ & $0^+_1 \rightarrow 0^+_1$ &  0.80 (0.09) & 0.75 (0.09) & 7$\%$ \\
~ & $0^+_1 \rightarrow 0^+_2$ & 0.74 (0.01) & 0.78 (0.001) & 5$\%$ \\
\midrule
  $^{76}$Ge  $\rightarrow$ $^{76}$Se & ~ & ~ & ~ & ~ \\
 ~ & $0^+_1 \rightarrow 0^+_1$ & 1.46 (0.20) & 1.59 (0.21) & 8$\%$ \\
 ~ & $0^+_1 \rightarrow 0^+_2$ & 0.10 (0.01) & 0.12 (0.02) & 20$\%$  \\
\midrule
  $^{82}$Se  $\rightarrow$ $^{82}$Kr & ~ & ~ & ~ & ~ \\
 ~ & $0^+_1 \rightarrow 0^+_1$ & 1.12 (0.13) & 1.26 (0.15) & 11$\%$ \\
 ~ & $0^+_1 \rightarrow 0^+_2$ & 0.17 (0.003) & 0.15 (0.00001) & 12$\%$ \\
\bottomrule
\end{tabular}
\end{table*}

As can be seen, the perturbative behavior represented in
Fig. \ref{48Ca_obo} is driven by the Gamow-Teller component, since the
short-range LO contact term is rather flat between second- and
third-order, and the perturbative expansion affects very weakly both
long-range Fermi and tensor matrix elements $M^{0\nu}_{\rm F}$,
$M^{0\nu}_{\rm T}$.

This is a feature that was already observed when analyzing the
perturbative behavior of the LO two-body currents.

Same observations are valid for the \zbb~ decay of the ground states
of $^{76}$Ge, $^{82}$Se into the one of $^{76}$Se, $^{82}$Kr, as can
be inferred from the inspection of Figs. \ref{76Ge_obo} and
\ref{82Se_obo}, where they are reported the calculated values of \nme,
$M^{0\nu}_{\rm GT}$, $M^{0\nu}_{\rm F}$, $M^{0\nu}_{\rm T}$, and
$M^{0\nu}_{\rm S}$ from first- up to third-order in perturbation
theory, as well as the value of their Pad\'e approximant $[2|1]$.

This exposition of our results leads to an analysis of our evaluated
\nmes~ for $^{48}$Ca, $^{76}$Ge, and $^{82}$Se \zbb~ decays, as well
as an estimate of the corresponding theoretical uncertainties.

As already pointed out in Ref. \cite{Coraggio26}, a calculation of
\nme~ that is grounded on many-body perturbation theory does not allow
to define a proper theoretical error, then we evaluate the uncertainty
affecting our predicted values by considering the theory of Pad\'e
approximant, that is a tool to obtain the best approximation of the
sum of a perturbative expansion from a truncated power series
\cite{Baker70}.

Then, we have reported in Table \ref{NME} the values of \nmes~ for the
decays of the  $J^{\pi}=0^+_1$ (ground state) of the parent nuclei
into the  $J^{\pi}=0^+_1$ and  $J^{\pi}=0^+_2$ states of the
grand-daughter nuclei, obtained with LO and N$^2$LO two-body currents
at third order in many-body perturbation theory, and also the
Pad\'e approximant $[2|1]$ which accounts for the perturbative
behavior up to third-order.
In the fifth and eighth columns, the absolute value of the difference
between the values $\hat{\Delta}=|M^{0\nu}_{\rm 3rd} - M^{0\nu}_{\textrm
  {Pad\'e}}|$ have been reported, which, as in our previous work
\cite{Coraggio26}, we consider the estimate of the uncertainties
associated to our calculated \nmes.

As can be observed in Table \ref{NMEobo}, the \nmes, as calculated at
LO and N$^2$LO, differ in most cases around 10$\%$, the largest
discrepancy being the one for the \nme~ of the ground-state decay of
$^{76}$Ge into the yrare $J^{\pi}=0^+$ of $^{76}$Se.
This evidences an excellent perturbative behavior with respect to the
ChPT expansion of the \zbb~ two-body currents, and leads to consider
the LO expression of the \zbb-decay operator as a sound approximation
to evaluate the nuclear matrix elements.

\begin{table*}[t]
\caption{Long-range LO matrix elements and corrections at N$^2$LO and
  N$^3$LO obtained through VMC and SM. At N$^2$LO, the VMC result
  includes only one three-body term generated by
  explicit $\Delta$ degrees of freedom. The N$^3$LO
  entries include only three-body diagrams. The last two columns give
  the corrections relative to the corresponding M$_{\mathrm{long,LO}}$.}
\label{tab:vmc_rsm_perturbative_pattern}
\setlength{\tabcolsep}{9pt}
\renewcommand{\arraystretch}{1.25}
\begin{tabular}{ccrrrrr}
\toprule
Method & $A$ & $M^{0\nu}_{\mathrm{long,LO}}$ & $M^{0\nu}_{\mathrm{N}^2\mathrm{LO}}$ & $M^{0\nu}_{\mathrm{N}^3\mathrm{LO},\,\mathrm{3B}}$ & $M_{\mathrm{N}^2\mathrm{LO}}/M_{\mathrm{long,LO}}$
& $M_{\mathrm{N}^3\mathrm{LO},\,\mathrm{3B}}/M_{\mathrm{long,LO}}$ \\
\midrule
VMC & 6  & $4.6915$ & $ 0.0117$ & $-0.0009$ & $ 0.25\%$ & $-0.19\,\permille$ \\
    & 8  & $0.1817$ & $-0.0228$ & $ 0.0049$ & $-13\%$    & $ 2.7\%$          \\
    & 12 & $0.8792$ & $-0.0588$ & $ 0.0011$ & $-6.7\%$    & $ 0.13\%$         \\
\midrule
SM & 6  & $5.0123$ & $ 0.3970$ & $ 0.0003$ & $ 7.9\%$   & $ 0.060\,\permille$ \\
    & 8  & $0.3223$ & $ 0.0179$ & $ 0.0077$ & $ 5.6\%$   & $ 2.4\%$           \\
    & 12 & $2.1023$ & $ 0.0807$ & $ 0.0086$ & $ 3.8\%$   & $ 0.41\%$          \\
\bottomrule
\end{tabular}
\end{table*}

\begin{table*}[t]
  \caption{Same as in last three columns of Table \ref{NME}, but
    including in the \zbb-decay operator also the contribution of
    three-body currents, calculated at N$^3$LO in ChPT.}
\label{NME3BC}
\setlength{\tabcolsep}{9pt}
\renewcommand{\arraystretch}{1.25}
\begin{tabular}{llccc}
\toprule
Decay & $J^{\pi}_i \rightarrow J^{\pi}_f$ &
$M^{0\nu}_{\rm 3rd}$ & $M^{0\nu}_{\textrm {Pad\'e}}$ & $\hat{\Delta}$ \\
\midrule
  $^{48}$Ca  $\rightarrow$ $^{48}$Ti & ~ & ~ & ~ & ~ \\
~ & $0^+_1 \rightarrow 0^+_1$ & 0.84 & 0.75 & 0.09 \\
~ & $0^+_1 \rightarrow 0^+_2$ & 0.78 & 0.77 & 0.01 \\
\midrule
  $^{76}$Ge  $\rightarrow$ $^{76}$Se & ~ & ~ & ~ & ~ \\
 ~ & $0^+_1 \rightarrow 0^+_1$ & 1.79 & 1.58 & 0.21 \\
 ~ & $0^+_1 \rightarrow 0^+_2$ & 0.10 & 0.12 & 0.02 \\
\midrule
  $^{82}$Se  $\rightarrow$ $^{82}$Kr & ~ & ~ & ~ & ~ \\
 ~ & $0^+_1 \rightarrow 0^+_1$ & 1.40 & 1.26 & 0.14 \\
 ~ & $0^+_1 \rightarrow 0^+_2$ & 0.15 & 0.15 & 0.00005 \\
\bottomrule
\end{tabular}
\end{table*}

We now shift the attention on the contribution of the three-body
component of the \zbb-decay operator, that has been obtained deriving
the three-body currents (3BC) at N$^3$LO in ChPT (see the details in
Sec. \ref{theory} and Appendix \ref{3BC}), and then performing a
normal-ordering procedure to construct a density-dependent
two-body component which is accounted for in the derivation of
\thetaeff.

First, it is worth reporting about a comparison of the perturbative
properties of the ChPT expansion of \zbb-decay current operators as
they emerge in fictitious decays in light nuclear systems, through
calculations performed by employing the variational Monte Carlo method
(VMC) Ref. \cite{Chambers-Wall26} and our shell-model approach.
More precisely we are going to consider the \zbb~ transitions for
$p$-shell nuclei $^{6}\mathrm{He}\rightarrow{}^{6}\mathrm{Be}$,
$^{8}\mathrm{He}\rightarrow{}^{8}\mathrm{Be}$, and
$^{12}\mathrm{Be}\rightarrow{}^{12}\mathrm{C}$.

It should be stressed that in Ref. \cite{Chambers-Wall26}, the
starting point was the Norfolk NV-Ia$^*$ ChPT nuclear Hamiltonian
\cite{Piarulli16}, while we employ the Entem-Machleidt N$^3$LO,
supplemented by a consistent three-body component (see Section
\ref{theory}).

Since the Norfolk interaction explicitly includes $\Delta$-isobar
degrees of freedom, whereas our nuclear Hamiltonian does not, a
term-by-term comparison of the individual contributions is not
possible, and the forecoming discussion has not to be considered a
benchmark calculation.
We want to just analyze whether or not the two different approaches
lead to consistent perturbative properties, with respect to the ChPT
expansion of the \zbb~ currents.

Therefore, we compare the results obtained with the two methods,
considering each correction relative to the long-range component of
the leading-order matrix element ($M^{0\nu}_{\rm long,LO}$).

Beyond leading order, the first term in the VMC calculation of
Ref.~\cite{Chambers-Wall26} is the one at N$^2$LO, which accounts only
for the three-body current induced by the explicit inclusion of the
$\Delta$ degrees of freedom, while the complete two-body correction is
not calculated.
It is worth noting that in our SM calculations, whose Hamiltonian does
not include the $\Delta$, N$^2$LO contributions are limited to those
characterizing the two-body current operator.
Then, we intend to compare only the order of magnitude of N$^2$LO
corrections with respect to the corresponding leading-order
$M^{0\nu}_{\rm long,LO}$.

In this context, it is worth noting that the ultrasoft contributions
to the two-body currents at N$^2$LO,  are omitted in the SM
calculations because of their explicit dependence on the decay
$Q_{\beta\beta}$ value, as shown in Eq.~ (\ref{eq:usoft}).
For the $^{6}$He decay, the negative $Q_{\beta\beta}$ value makes the
decay energetically forbidden.
For the $^{8}$He and $^{12}$Be decays, the experimental
$Q_{\beta\beta}$ values are approximately 27 and 25~MeV, respectively,
substantially larger than the approximately 2--4~MeV typical of the
\zbb~candidates.
As a consequence, Eq.~(\ref{eq:usoft}) yields unrealistic ultrasoft
contributions for $^{8}$He and $^{12}$Be.

Finally, at N$^3$LO we compare the three-body contributions only.
For all these reasons, the comparison is meant only as a qualitative
consistency check of the perturbative pattern.

\begin{table*}[t]
\caption{Same as in Table \ref{NME}, but employing a non-local
  regulator for the short-range \nmess and including also the
  contribution of three-body currents (3BC), calculated at N$^3$LO in
  ChPT.}
\label{NMEnlr}
\setlength{\tabcolsep}{9pt}
\renewcommand{\arraystretch}{1.25}
\begin{tabular}{llccc|ccc}
\toprule
~ & ~ & \multicolumn{3}{c}{LO} & \multicolumn{3}{c}{N$^2$LO plus 3BC} \\
Decay & $J^{\pi}_i \rightarrow J^{\pi}_f$ &
$M^{0\nu}_{\rm 3rd}$ & $M^{0\nu}_{\textrm {Pad\'e}}$ & $\hat{\Delta}$ &
$M^{0\nu}_{\rm 3rd}$ & $M^{0\nu}_{\textrm {Pad\'e}}$ & $\hat{\Delta}$ \\
\midrule
  $^{48}$Ca  $\rightarrow$ $^{48}$Ti & ~ & ~ & ~ & ~ & ~ & ~ & ~ \\
~ & $0^+_1 \rightarrow 0^+_1$ &  0.96 & 0.87 & 0.09 & 0.91 & 0.82 & 0.09 \\
~ & $0^+_1 \rightarrow 0^+_2$ & 0.79 & 0.78 & 0.01 & 0.82 & 0.81 & 0.01 \\
\midrule
  $^{76}$Ge  $\rightarrow$ $^{76}$Se & ~ & ~ & ~ & ~ & ~ & ~ & ~ \\
 ~ & $0^+_1 \rightarrow 0^+_1$ & 1.73 & 1.51 & 0.22& 1.86 & 1.64 & 0.22 \\
 ~ & $0^+_1 \rightarrow 0^+_2$ & 0.08 & 0.10 & 0.02 & 0.10 & 0.11 & 0.01 \\
\midrule
  $^{82}$Se  $\rightarrow$ $^{82}$Kr & ~ & ~ & ~ & ~ & ~ & ~ & ~ \\
 ~ & $0^+_1 \rightarrow 0^+_1$ & 1.31 & 1.17 & 0.14 & 1.46 & 1.31 & 0.15 \\
 ~ & $0^+_1 \rightarrow 0^+_2$ & 0.18 & 0.18 & 0.004 & 0.16 & 0.16 & 0.0001 \\
\bottomrule
\end{tabular}
\end{table*}

Within these limitations, the LO matrix elements are of comparable
magnitude in the two approaches, with particularly good agreement for
$A=6$.
More importantly, VMC and SM calculations display a broadly consistent
perturbative hierarchy, with the N$^2$LO contributions being smaller
than the corresponding LO matrix elements and the three-body terms at
N$^3$LO being further suppressed, although the size of the suppression
varies with the mass number.
With respect to the LO contribution, the available N$^2$LO ones range
from effects of less than $1\%$ to corrections of order $10\%$.
The N$^3$LO contributions are negligible for $A=6$, reach a few
percent for $A=8$, and remain below the percent level for $A=12$.

Thus, despite the differences in the nuclear interactions and in the
contributions included, both the VMC and SM calculations exhibit a
qualitatively similar perturbative hierarchy. 
In particular, both indicate that the three-body terms at N$^3$LO
remain subleading.

Let us now to meet such an issue for the nuclei under our current
investigation. 

In Table \ref{NME3BC}, the values of \nmes~ for the \zbb-decay of
$^{48}$Ca, $^{76}$Ge, and $^{82}$Se -- calculated with a
decay operator which includes the contribution of three-body currents
derived at N$^3$LO -- are reported.
Comparing the results of the sixth column of Table \ref{NME} with
those in the third column of Table\ref{NME3BC}, we may observe that
the contribution of three-body currents to the \nmes~ is less than
$1\%$, a result that is consistent with those that have been obtained
for the fictitious decay of $p$-shell nuclei with our SM approach.

The result in a $\Delta$-less approach we have obtained with the
inclusion of three-body currents are quite encouraging, since they
evidence that the leading contribution to \nmes~ comes from two-body
currents, three-body currents playing quite a minor role.

As mentioned before, we may safely conclude this part of the
discussion of our results that the main role in the calculated \nmes~
is played by the leading-role of the ChPT expansion of the \zbb-decay
currents.

We move now to another topic that has been anticipated in
Sec. \ref{intro} and \ref{theory}, namely the choice of the regulator
which characterizes the short-range LO component of the \zbb-decay
operator, and the value of the LEC $g_{\nu}^{\rm NN}$ which
regularizes such a term.

To this end, we have also performed the calculation of the \nmes~
employing a non-local regulator for the short-range component of the
nuclear matrix element \nmess, with a cutoff placed at $\Lambda=500$
MeV, namely a structure that is fully consistent with the one that is
considered both for the 2NF and 3NF components of the nuclear
Hamiltonian.

In Table \ref{NMEnlr} we have reported the values of the \nmes~
calculated with a non-local regulator for the short-range \nmess, and
using $g_{\nu}^{\rm NN}=-0.597$ fm$^2$ \cite{Wirth21}, with a
long-range \zbb-operator at LO and N$^2$LO, the latter case including
also the contribution of 3BC.

Comparing the results in Table \ref{NME3BC} and those
in Table \ref{NMEnlr} of this work, we observe that the values of the
\nmes, calculated with the two different approaches to the choice of
the regulator function for $\Theta_{\rm S}$ and $g_{\nu}^{\rm NN}$,
are within the boundaries of the uncertainties $\Delta$s.
We believe that it is important to underline this result, that support
the soundness of the procedures in
Refs. \cite{Cirigliano19,Jokiniemi21,Wirth21} to renormalize the LO
short-range component of the \zbb-decay operator.

Finally, we summarize the results of our study by attempting to
estimate also the uncertainty associated with the truncation of the
ChPT expansion of the \zbb-decay operator.
We will refer to the results reported at LO and N$^2$LO in Table
\ref{NMEnlr}, since we consider those numbers more consistent with our
choice of the chiral nuclear Hamiltonian, that is constructed in terms
of a non-local regulator.

Although these results also include the partial N$^3$LO contribution
due to the inclusion of 3BC in the \zbb-decay operator, its numerical
impact is negligible at the precision quoted here, and we therefore
use these values to represent the N$^2$LO matrix elements.
As a matter of fact, the N$^3$LO 3BC terms that are included in the
present calculation do not define a complete N$^3$LO decay operator.
They rather provide an additional diagnostic of the convergence
pattern of the  chiral expansion.
In fact, as observed before, their impact is found to be below the
percent level for the nuclei considered here, supporting the
conclusion that the dominant chiral effects are already accounted for
by the LO and N$^2$LO two-body currents.

\begin{table*}[t]
\caption{Estimate of the chiral truncation uncertainty obtained from
  the LO and N$^2$LO Padé matrix elements of Table \ref{NMEnlr}. We use
  $\Lambda=500$ MeV and $Q=0.4$.}
\label{chiral_error_nonlocal}
\setlength{\tabcolsep}{9pt}
\renewcommand{\arraystretch}{1.25}
\begin{tabular}{llccccc}
\toprule
Decay & $J^{\pi}_i \rightarrow J^{\pi}_f$ & \(M^{0\nu}_{\rm LO}\) &
\(M^{0\nu}_{\rm N2LO}\) & \(|\hat{\Delta}_{\rm LO-N2LO}|\)
& \(Q^3 |M^{0\nu}_{\rm LO}|\) & \(\hat{\Delta}_{\chi}^{>{\rm N2LO}}\) \\
\midrule
  $^{48}$Ca  $\rightarrow$ $^{48}$Ti & ~ & ~ & ~ & ~ & ~ & ~ \\
~ & $0^+_1 \rightarrow 0^+_1$ & 0.87 & 0.82 & 0.05 & 0.06 & 0.06 \\
~ & $0^+_1 \rightarrow 0^+_2$ & 0.78 & 0.81 & 0.03 & 0.05 & 0.05 \\
\midrule
  $^{76}$Ge  $\rightarrow$ $^{76}$Se & ~ & ~ & ~ & ~ & ~ & ~ \\
 ~ & $0^+_1 \rightarrow 0^+_1$ & 1.51 & 1.64 & 0.13 & 0.10 & 0.10 \\
 ~ & $0^+_1 \rightarrow 0^+_2$ & 0.10 & 0.11 & 0.01 & 0.01 & 0.01 \\
\midrule
  $^{82}$Se  $\rightarrow$ $^{82}$Kr & ~ & ~ & ~ & ~ & ~ & ~ \\
 ~ & $0^+_1 \rightarrow 0^+_1$ & 1.17 & 1.31 & 0.14 & 0.07 & 0.07 \\
 ~ & $0^+_1 \rightarrow 0^+_2$ & 0.18 & 0.16 & 0.02 & 0.01 & 0.01 \\
\bottomrule
\end{tabular}
\end{table*}

\begin{table*}[t]
\caption{Summary of the theoretical uncertainties associated with the
  non-local-regulator results of Table \ref{NMEnlr}. The central
  values are the Pad\'e N$^2$LO matrix elements.}
\label{total_uncertainty}
\setlength{\tabcolsep}{9pt}
\renewcommand{\arraystretch}{1.25}
\begin{tabular}{llcccc}
\toprule
Decay & $J^{\pi}_i \rightarrow J^{\pi}_f$ & \(M^{0\nu}_{\rm Pad\acute e}\)
& \(\hat{\Delta}_{\rm MBPT}\) & \(\hat{\Delta}_{\chi}\) & \(\hat{\Delta}_{\rm tot}\) \\
\midrule
  $^{48}$Ca  $\rightarrow$ $^{48}$Ti & ~ & ~ & ~ & ~ & ~ \\
~ & $0^+_1 \rightarrow 0^+_1$ & 0.82 & 0.09 & 0.06 & 0.15 \\
~ & $0^+_1 \rightarrow 0^+_2$ & 0.81 & 0.01 & 0.05 & 0.06 \\
\midrule
  $^{76}$Ge  $\rightarrow$ $^{76}$Se & ~ & ~ & ~ & ~ & ~ \\
 ~ & $0^+_1 \rightarrow 0^+_1$ & 1.64 & 0.22 & 0.10 & 0.32 \\
 ~ & $0^+_1 \rightarrow 0^+_2$ & 0.11 & 0.01 & 0.01 & 0.02 \\
\midrule
  $^{82}$Se  $\rightarrow$ $^{82}$Kr & ~ & ~ & ~ & ~ & ~ \\
 ~ & $0^+_1 \rightarrow 0^+_1$ & 1.31 & 0.15 & 0.07 & 0.22 \\
 ~ & $0^+_1 \rightarrow 0^+_2$ & 0.16 & 0.0001 & 0.01 & 0.01 \\
\bottomrule
\end{tabular}
\end{table*}

Following the order-by-order prescription proposed by
Epelbaum, Krebs, and Mei{\ss}ner~\cite{Epelbaum14}, adapted to the
$\Delta$-less counting of the \zbb-decay operator, we estimate the
residual chiral truncation uncertainty as

\[
\hat{\Delta}_{\chi}^{>{\rm N2LO}} = \max\left[ Q^3\left|M^{0\nu}_{\rm LO}\right|,
Q\left|M^{0\nu}_{\rm N2LO} - M^{0\nu}_{\rm LO}\right|
\right] ~.
\]

Taking a breakdown scale $\Lambda=500$ MeV and a typical
virtual-neutrino momentum $p_{\rm typ}\simeq 200$ MeV gives
$Q=0.4$.
Using the Pad\'e values of Table \ref{NMEnlr}, we obtain the values of
$\hat{\Delta}_{\chi}^{>{\rm N2LO}}$ that are reported in Table
\ref{chiral_error_nonlocal} for the decays of $^{48}$Ca, $^{76}$Ge, and
$^{82}$Se, from the ground state of the parent nuclei into the yrast
and yrare $J^{\pi}=0^+$ states of the grand-daughter ones.
These correspond to relative uncertainties not exceeding $10\%$ with
respect to the N2LO matrix elements.

We conclude commenting on the overall theoretical uncertainty
associated with the calculated nuclear matrix elements.
The chiral truncation uncertainty discussed above probes the
convergence of the $\Delta$-less chiral expansion of the \zbb-decay
operator.
It is therefore distinct from the uncertainty associated
with the many-body perturbative derivation of the effective
shell-model operator.
The latter is estimated, as in the previous tables, from the absolute
difference $\hat{\Delta}_{\rm MBPT}$ between the third-order result
and the corresponding Pad\'e approximant, 

\[
\hat{\Delta}_{\rm MBPT} = \left| M^{0\nu}_{\rm 3rd} - M^{0\nu}_{\rm Pad\acute e}
\right| ~.
\]

Using the non-local-regulator results of Tables \ref{NMEnlr} and
\ref{chiral_error_nonlocal}, we summarize the values of the
uncertainties in the last three columns of Table \ref{total_uncertainty}.

The uncertainty reported in the fourth column reflects the
perturbative derivation of the effective SM operator \thetaeff,
whereas the one inserted in the fifth column estimates the residual
truncation error of the chiral \zbb-decay operator.

Since both uncertainties have a systematic character, a single
conservative uncertainty estimate can be obtained combining them linearly,

\[
\hat{\Delta}_{\rm tot} = \hat{\Delta}_{\rm MBPT} + \hat{\Delta}_{\chi} ~,
\]

\noindent
which is reported in the last column of Table \ref{total_uncertainty}
for the \zbb~ decays of $^{48}$Ca, $^{76}$Ge, and $^{82}$Se decays.
These correspond to relative uncertainties of about $20\%$ for the
decays to the ground states of grand-daughter nuclei, and between $6\%$
and $20\%$ for those to the yrare $J^{\pi}=0^+$.

Other possible sources of theoretical uncertainty, such as the
residual dependence on the nuclear Hamiltonian, the shell-model
valence space, and the values of the short-range low-energy constants,
are not folded into the above estimate and should be assessed
separately.

\section{Summary and conclusions}\label{conclusions}

This paper has been devoted to a thorough analysis of the theoretical
aspects which characterize the calculation of \zbb-decay nuclear
matrix elements when starting from the perspective of effective field
theory, whose goal is to account for QCD symmetries at the low-energy
scale of nuclear systems. 
More precisely, the building blocks of our investigation have been
nuclear Hamiltonians and decay operators that have been constructed
through a chiral perturbative expansion, and then employed to
calculate nuclear wave functions and matrix elements within the shell
model framework.

This has implied the derivation of effective SM Hamiltonians and
\zbb-decay operators to account for the configurations of the nuclear
Hamiltonian that are not explicitly included in the truncated model
space, a task that has been pursued by way of many-body perturbation
theory.

Then, the main issue to be tackled within such an approach is to
verify the robustness of the predicted \nmes~ with respect to the ChPT
expansion of the operators and of the effective SM operators, and try
to quantify the uncertainties originated by the perturbative
procedures.

It is worth mentioning that we consider present work as the new
chapter of a systematic study of electroweak nuclear-decay processes
in intermediate-mass nuclei ($^{48}$Ca, $^{76}$Ge, $^{82}$Se)
which has started by benchmarking our EFT-based approach with GT
observables -- as well as low-energy spectroscopic data
\cite{Coraggio24a}--, then moving to the prediction of \zbb-decay
nuclear matrix elements with operators derived at leading order in
ChPT \cite{Coraggio26}.

As reported in Sec. \ref{results}, we have met a few subjects which
may be highlighted as follows:

\begin{itemize}
\item as in our previous study \cite{Coraggio26}, we have estimated
  the uncertainty that is linked to the many-body perturbative
  expansion of the effective SM operator \thetaeff, by way of the
  theory of Pad\'e approximant \cite{Baker70}, which affects the
  predicted \nmes~ for the decay of the ground state of the parent
  nucleus into the one of the grand-daughter system utmost $15 \%$.
  The uncertainty related to the \nmes~ of the decay to the yrare
  $J^{\pi}=0^+$ of the grand-daughter nuclei are generally
  smaller, being less than $10 \%$.
\item The uncertainties of our predicted \nmes, and related to the
  ChPT expansion of the \zbb-decay operator are smaller, not exceeding
  $7 \%$.
  This result is consistent with another similar investigation in
  Ref. \cite{Castillo25}, which was carried out in terms of
  phenomenological nuclear Hamiltonians, and establishes that at
  N$^2$LO of the ChPT expansion of two-body currents we
  obtain a sound definition of the \zbb-decay operator.
\item The impact of the contribution of three-body currents to the
  \zbb-decay operator, which in our $\Delta$-less approach appears at
  N$^3$LO, is almost negligible, contributing for a few percent to the
  calculated \nmes.
\item We have tested two possible approaches to the renormalization of
  the LO contact term, which appears in the expression of the \zbb~
  two-body currents.
  Both approaches fix the LEC $g_{\nu}^{\rm NN}$, that characterize
  this contact term, by considering the $nn \rightarrow pp e^- e^-$
  process and employing the 2NF component of the Entem-Machleidt
  chiral Hamiltonian, but considering local and non-local regulator
  functions, respectively \cite{Cirigliano19,Jokiniemi21,Wirth21}.
  The results in Tables \ref{NME3BC} and \ref{NMEnlr} evidence that
  the choice of a local or non-local regulator to calculate LO short-range
  \nmess~ does not lead to substantially different \nmes.
\item The \nmes~ for the decay from parent-nucleus ground states to
  grand-daughter-nucleus yrare $J^{\pi}=0^+$ are about one order of
  magnitude smaller with respect to those for the decay into the
  ground state of grand-daughter nuclei for $^{76}$Ge and $^{82}$Se,
  making these decay processes quite unlikely to be observed.
  The analogous \nmes~ for the $^{48}$Ca decay are almost equal, but
  it should be pointed out that decay to yrare $^{48}$Ti $J^{\pi}=0^+$
  is suppressed with respect to the one to the ground state, because
  its $Q_{\beta \beta}$ value is 3 MeV smaller.
\end{itemize}

In relation to our estimate of the theoretical uncertainty of the
\nmes~ reported in this paper, it should be mentioned the work of
Ref. \cite{Belley24}, where the authors carried out a comprehensive
{\it ab initio} uncertainty quantification of the \zbb~ decay of
$^{76}$Ge, employing also nuclear Hamiltonians and electroweak
operators derived within ChPT.
There, $^{76}$Ge \nme~ was calculated with recently developed
many-body emulators, leading to a numerical result $M^{0\nu} =
2.60^{+1.28}_{-1.36}$, which is consistent within the theoretical
error with our value in Table \ref{total_uncertainty}, namely
$M^{0\nu} =1.64 (0.32)$.

Next chapters of our project will be to perform similar analyses for
the decay of heavier systems, $^{100}$Mo, $^{130}$Te, and $^{136}$Xe,
which will be in a near future the target of future ton-scale
experiments \cite{Armengaud21,Adhikari21}, and also perform an
extended comparison of results obtained with SM and {\it ab initio}
calculations, in order to proceed towards benchmark procedures which
could validate many-body approaches for \zbb-decays of large-mass
nuclei.

\section{Acknowledgments}
The calculations were made possible by grants of computing time from
the Italian National Supercomputing Center CINECA, under the
CINECA-INFN agreement.
The authors thank Dr Saori Pastore, Dr Graham Chambers-Wall and Dr
Emanuele Mereghetti for useful discussion on the ChPT expansion of
neutrinoless double-$beta$ decay current operators.
\appendix

\section{N$^2$LO neutrino potentials}
\label{app:neutrino-potentials}

For the $0^+\!\to0^+$ transitions considered in this work, only the
parity-even contributions generated by two insertions of either the
vector current (VV) or the axial current (AA) are required.
The N$^2$LO two-body neutrino potential is written as
\begin{equation}
\begin{aligned}
V_{\nu,2}^{(m,n)}
={}&
\tau_m^-\tau_n^-
\Bigg[
    \mathcal V_{VV}^{(m,n)}
    +\mathcal V_{AA}^{(m,n)}
\\
&\quad
    +\widetilde{\mathcal V}_{AA}^{(m,n)}
     \ln\left(\frac{m_\pi^2}{\mu_{\mathrm{us}}^2}\right)
    +\mathcal V_{\mathrm{CT}}^{(m,n)}
\Bigg].
\end{aligned}
\label{eq:n2lo-potential}
\end{equation}
Each term is decomposed into Fermi, Gamow-Teller, and tensor
components according to
\begin{equation}
\begin{aligned}
\mathcal V_X^{(m,n)}
={}&
\mathcal V_X^{\rm F}
+\mathcal V_X^{\rm GT}\,
 \bm{\sigma}_m\!\cdot\!\bm{\sigma}_n
\\
&+
\mathcal V_X^{\rm T}\,
S_{mn}(\hat{\bm q}),
\end{aligned}
\label{eq:potential-decomposition}
\end{equation}
where
\begin{equation}
S_{mn}(\hat{\bm q})
=
\bm{\sigma}_m\!\cdot\!\bm{\sigma}_n
-
3
\frac{
(\bm{\sigma}_m\!\cdot\!\bm q)
(\bm{\sigma}_n\!\cdot\!\bm q)
}{
\bm q^{\,2}
}.
\label{eq:tensor-operator}
\end{equation}
In the following, the particle labels $(m,n)$ are suppressed.
We introduce the dimensionless momentum variable and the chiral
logarithm
\begin{equation}
    \hat q\equiv\frac{\bm q^{\,2}}{m_\pi^2},
    \qquad
    L_\pi\equiv
    \ln\left(\frac{\mu^2}{m_\pi^2}\right).
\label{eq:qhat-Lpi}
\end{equation}
Moreover, we define the loop function as
\begin{equation}
    g(\hat q)
    =
    \frac{4}{\sqrt{\hat q(4+\hat q)}}
    \operatorname{arctanh}
    \left(
        \sqrt{\frac{\hat q}{4+\hat q}}
    \right),
\label{eq:loop-function}
\end{equation}
\noindent
and the auxiliary functions 
\begin{equation}
\begin{aligned}
f_0(\hat q)
={}&
-\frac{1+8\hat q}{6\hat q}
+\frac{(1+\hat q)(1+8\hat q+\hat q^2)}
       {6\hat q^2}
 \ln(1+\hat q)
\\
&-
\frac{(4+\hat q)(5+2\hat q)}{24}
g(\hat q),
\end{aligned}
\label{eq:f0}
\end{equation}
\begin{equation}
\begin{aligned}
f_2(\hat q)
={}&
\frac{1+8\hat q}{3\hat q}
+\frac{(1+\hat q)^2(-1+5\hat q)}
       {3\hat q^2}
 \ln(1+\hat q)
\\
&-
\frac{40+47\hat q+10\hat q^2}{12}
g(\hat q),
\end{aligned}
\label{eq:f2}
\end{equation}
\begin{equation}
\begin{aligned}
f_4(\hat q)
={}&
-\frac{1}{6}
\left(
    20+\frac{1}{\hat q}
    -\frac{12}{4+\hat q}
\right)
\\
&-
\frac{
    -1+14\hat q+78\hat q^2
    +62\hat q^3+23\hat q^4
}{
    6\hat q^2(1+\hat q)
}
\ln(1+\hat q)
\\
&+
\frac{
    640+912\hat q+375\hat q^2+46\hat q^3
}{
    24(4+\hat q)
}
g(\hat q).
\end{aligned}
\label{eq:f4}
\end{equation}

\paragraph{Vector contribution}
The non-vanishing vector-vector components are
\begin{align}
\mathcal V_{VV}^{\rm GT}
&=
-\frac{\hat q}{3}
 \frac{g_A^2}{(4\pi f_\pi)^2}
 \mathcal F_{VV}(\hat q),
\label{eq:VV-GT}
\\
\mathcal V_{VV}^{\rm T}
&=
+\frac{\hat q}{3}
 \frac{g_A^2}{(4\pi f_\pi)^2}
 \mathcal F_{VV}(\hat q),
\label{eq:VV-T}
\end{align}
where
\begin{equation}
\begin{aligned}
\mathcal F_{VV}(\hat q)
={}&
\frac{2(1-\hat q)^2}
     {\hat q^2(1+\hat q)}
\ln(1+\hat q)
-\frac{2}{\hat q}
\\
&+
\frac{7-3\hat q L_\pi}
     {(1+\hat q)^2}
+\frac{L_\pi}{1+\hat q}.
\end{aligned}
\label{eq:F-VV}
\end{equation}

\paragraph{Axial contribution}

The Fermi component of the axial part is
\begin{equation}
\begin{aligned}
\mathcal V_{AA}^{\rm F}
={}&
\frac{1}{(4\pi f_\pi)^2}
\Bigg[
-\frac{3}{4}(1-g_A^2)^2L_\pi
\\
&\quad
+g_A^4f_4(\hat q)
+g_A^2f_2(\hat q)
+f_0(\hat q)
\\
&\quad
+24g_A^2f_\pi^2C_T(L_\pi+1)
\Bigg].
\end{aligned}
\label{eq:AA-F}
\end{equation}
The corresponding Gamow-Teller and tensor components are
\begin{align}
\mathcal V_{AA}^{\rm GT}
&=
+\frac{\hat q}{3}
 \frac{g_A^2}{(4\pi f_\pi)^2}
 \mathcal F_{AA}(\hat q),
\label{eq:AA-GT}
\\
\mathcal V_{AA}^{\rm T}
&=
-\frac{\hat q}{3}
 \frac{g_A^2}{(4\pi f_\pi)^2}
 \mathcal F_{AA}(\hat q),
\label{eq:AA-T}
\end{align}
where
\begin{equation}
\mathcal F_{AA}(\hat q)
=
\frac{g_A^2}{1+\hat q}(L_\pi-4)
+\frac{1}{(1+\hat q)^2}.
\label{eq:F-AA}
\end{equation}

\paragraph{Loop-Ultrasoft contribution}

The coefficients multiplying the loop-ultrasoft term
Eq. (\ref{eq:n2lo-potential}) are
\begin{equation}
\begin{aligned}
\widetilde{\mathcal V}_{AA}^{\rm F}
={}&
2\frac{g_A^4}{(4\pi f_\pi)^2}
\frac{\bm q^{\,2}}
     {\bm q^{\,2}+m_\pi^2}
-\frac{48g_A^2C_T}{(4\pi)^2},
\\
\widetilde{\mathcal V}_{AA}^{\rm GT}
={}&
+\frac{2}{3}
\frac{g_A^4}{(4\pi f_\pi)^2}
\frac{\bm q^{\,2}}
     {\bm q^{\,2}+m_\pi^2},
\\
\widetilde{\mathcal V}_{AA}^{\rm T}
={}&
-\frac{2}{3}
\frac{g_A^4}{(4\pi f_\pi)^2}
\frac{\bm q^{\,2}}
     {\bm q^{\,2}+m_\pi^2}.
\end{aligned}
\label{eq:AA-log-components}
\end{equation}

\paragraph{Counterterm contribution}

The Fermi counterterm contribution is
\begin{equation}
\mathcal V_{\mathrm{CT}}^{\rm F}
=
-\frac{2g_\nu^{\rm NN}}{(4\pi f_\pi)^2}.
\label{eq:CT-F}
\end{equation}
The Gamow-Teller and tensor contributions are
\begin{align}
\mathcal V_{\mathrm{CT}}^{\rm GT}
&=
+\frac{\hat q}{3}
 \frac{g_A^2}{(4\pi f_\pi)^2}
 \mathcal F_{\mathrm{CT}}(\hat q),
\label{eq:CT-GT}
\\
\mathcal V_{\mathrm{CT}}^{\rm T}
&=
-\frac{\hat q}{3}
 \frac{g_A^2}{(4\pi f_\pi)^2}
 \mathcal F_{\mathrm{CT}}(\hat q),
\label{eq:CT-T}
\end{align}
where
\begin{equation}
\mathcal F_{\mathrm{CT}}(\hat q)
=
\frac{5}{6}g_\nu^{\pi\pi}
\frac{\hat q}{(1+\hat q)^2}
-
g_\nu^{\pi {\rm N}}
\frac{1}{1+\hat q}.
\label{eq:F-CT}
\end{equation}
The renormalization-scale dependence of $g_\nu^{\pi\pi}$, $g_\nu^{\pi
  {\rm N}}$, and $g_\nu^{\rm NN}$ cancels the corresponding $\mu$
dependence of the vector-vector and axial-axial loop contributions. 
In our calculation, we chose two renormalization scales, $\mu=500$ MeV
and $\mu_\mathrm{us}=m_\pi$, respectively. In this case, the
contribution from $\widetilde{\mathcal V}_{AA}$ is eliminated. The
value of $C_T$ is taken to be the same as that used in the chiral
Hamiltonian.
For the three contact LECs, we use
$g_\nu^{\pi\pi}=-10.84-(36/5)\ln(\mu/m_\rho)$, taken from the average
of two lattice QCD calculations~\cite{Tuo19,Detmold20},
$g_\nu^{\pi {\rm N}}=0$, and $g_\nu^{\rm NN}=-0.47,-0.597$ fm${}^2$,
when considering a local or non-local regulator for the short-range
contact term, respectively (see comments in Sec. \ref{theory}).

\subsection{\zbb~ three-body neutrino potentials}
\label{3BC}

In chiral power counting without explicit $\Delta$ degrees of freedom,
the 3N neutrino potentials first appear at \NNNLO~,
in momentum space they are given by
\begin{widetext}
\begin{align}
  V_{\kappa_1} &= \frac{g_A^2}{4m_N f_\pi^2}(1+\kappa_1) \sum_{i\neq j\neq k}
    \frac{(\vsigma_j\cdot\vec{q}_j)(\vsigma_k\cdot\vec{q}_k)}{\vec{q}_i^2
    (\vec{q}_j^2+m_\pi^2)(\vec{q}_k^2+m_\pi^2)}
    \vsigma_i \cdot (\vec{q}_i\times\vec{q}_j) \tau_i^- [\vtau_k \times\vtau_j]^-  \,, \\
  V_{c_1} &= \frac{4g_A^2m_\pi^4c_1}{f_\pi^2}\sum_{i\neq j\neq k}
    \frac{(\vsigma_k\cdot\vec{q}_k)(\vsigma_i\cdot\vec{q}_i)}
    {\vec{q}_i^2(\vec{q}_k^2+m_\pi^2)(\vec{q}_i^2+m_\pi^2)^2} \tau_i^-\tau_k^- \,,\\
  V_{c_3} &= \frac{2g_A^2c_3}{f_\pi^2}\sum_{i\neq j\neq k}
    \frac{\vsigma_k\cdot\vec{q}_k}{\vec{q}_i^2(\vec{q}_k^2+m_\pi^2)}
    \left(\vsigma_i\cdot\vec{q}_k-(\vsigma_i\cdot\vec{q}_i)(\vec{q}_i\cdot\vec{q}_k)
    \frac{\vec{q}_i^2+2m_\pi^2}{(\vec{q}_i^2+m_\pi^2)^2}\right)\tau_i^-\tau_k^- \,, \\
  V_{c_4} &= \frac{g_A^2c_4}{f_\pi^2} \sum_{i\neq j\neq k}
    \frac{\vsigma_k\cdot\vec{q}_k}{\vec{q}_i^2(\vec{q}_k^2+m_\pi^2)}
    \left(-\vec{q}_k \cdot(\vsigma_i\times\vsigma_j) - (\vsigma_i\cdot\vec{q}_i)\vsigma_j
    \cdot(\vec{q}_i\times\vec{q}_k)\frac{\vec{q}_i^2+2m_\pi^2}{(\vec{q}_i^2+m_\pi^2)^2}\right)
    \tau_i^-(\vtau_k\times\vtau_j)^- \,, \\
  V_{c_d} &= -\frac{g_Ac_D}{2\Lambda_\chi f_\pi^2} \sum_{i\neq j\neq k}
    \frac1{\vec{q}_i^2}\left(\vsigma_i\cdot\vsigma_k - (\vsigma_i\cdot\vec{q}_i)(\vsigma_k\cdot\vec{q}_i)
    \frac{\vec{q}_i^2+2m_\pi^2}{(\vec{q}_i^2+m_\pi^2)^2}\right) \tau_i^-\tau_k^- \,.
\end{align}
\end{widetext}

\subsection{Matrix elements of \zbb~ three-body currents}

In this section, we describe the calculation of the three-body matrix
elements of the N$^3$LO three-nucleon neutrino potentials between
antisymmetrized three-nucleon harmonic oscillator (HO) basis states.
The overall procedure closely follows that employed for three-nucleon
force (3NF) matrix elements, and further details can be found in
Refs. \cite{Navratil07b,Roth14,Fukui18,Miyagi23}.

The starting point is the operator representation of the neutrino
potentials given in Sec. \ref{theory}.
To provide an illustrative example, the derivation of the $V_{c_1}$
contribution is presented here, with analogous treatment of the
remaining terms.
Exploiting the permutation symmetry of the three nucleons, we can
define a reduced operator as a function of the transferred momenta
$\vec{q}_i$.
For one single particle ordering configuration, the operator assumes
the following form:

\begin{multline}
  w_{c_1} \left(\vec{q}_a,\vec{q}_b,\vec{q}_c\right) \equiv
  \frac{1}{(2\pi)^6} \frac{8g_{A}^{2} m_\pi^4 c_1}{f_\pi^{2}}
  \tau_{b}^- \tau_{c}^- \\
  \null \times \frac{(\vsigma_{b}\cdot\vec{q}_b)
    (\vsigma_{c}\cdot\vec{q}_c)}{\vec{q}_c^2 (\vec{q}_b^2 + m_\pi^2)
    (\vec{q}_c^2 + m_\pi^2)^2} \,,
\end{multline}

\noindent
where $\sigma_i (\tau_i)$ denotes the spin (isospin) Pauli matrix of
nucleon $i$.

The first step in this procedure is to evaluate the matrix elements
(MEs) of the neutrino potentials in the partial-wave Jacobi-HO basis,
which is defined as

\begin{equation}
  \left|\bar{\alpha}; JT \right\ra \equiv |\bar{n}_{12} \bar{n};
  [(\bar{l}_{12}\bar{S}_{12}) \bar{I}_{12}, (\bar{l}\frac12) \bar{I}]
  J; (\bar{T}_{12}\frac12)T \ra\,,
\end{equation}

\noindent
and the quantum numbers with and without the subscript ``12'' are
associated with the first and second Jacobi coordinate, respectively:

\begin{equation}
  \vec{k} = \frac{1}{\sqrt{2}} (\vec{p}_a - \vec{p}_b)\,, \quad
  \vec{K} = \sqrt{\frac23}\left[ \frac12(\vec{p}_a + \vec{p}_b) -
    \vec{p}_c \right] \,.
\end{equation}

The collective index $\bar{\alpha}$ comprises all quantum numbers,
with the exception of the total three-nucleon angular momentum $J$ and
the total isospin $T$.
The principal quantum number $\bar{n}$, orbital angular momentum
$\bar{l}$, spin $\bar{S}$, and the total angular momentum $\bar{I}$
with respect to the two Jacobi momenta are all components of the
collective index.

To regularize the neutrino potential, we introduce a nonlocal Gaussian
regulator,

\begin{equation}
  f_\text{reg} (k,K;\Lambda_0) = \exp\left[ -\left(\frac{k^2 +
        K^2}{\Lambda_0^2}\right)^2 \right] \,.
\end{equation}

\noindent
The momentum cutoff is chosen as $\Lambda_0=500$ MeV, consistent with
that adopted for both the two-body neutrino potential and the nuclear
force.
The regularized potential can then be expressed in terms of the four
momentum variables $k, k', K$ and $K'$, together with the three
relative angles between $(\vec{k}, \vec{k}')$, $(\vec{K}, \vec{K}')$
and $(\vec{K}-\vec{K}', \vec{k}-\vec{k}')$.
After expanding the angular dependence in terms of multiple
partial-wave components, the regularized matrix element for the $c_1$
term can be written as

\begin{widetext}
\begin{multline}
    \la \bar{\alpha}'; J T^\prime|w_{c_1} | \bar{\alpha}; J T \ra =
    - \frac{54\sqrt{15}}{\pi^4} c_1 m_\pi^4 \frac{g_A^2 }{f_\pi^2} \\
    \null \times S^{JT}_{\bar{\alpha}\bar{\alpha}^\prime}
    \SixJ{\frac12}{\frac12}{\bar{T}_{12}}{\bar{T}_{12}^{\prime}}{1}{\frac12}
      \SixJ{\bar{S}_{12}}{\bar{S}_{12}^{\prime}}{1}{\frac12}{\frac12}{\frac12}
    \sum_{L_1 L_2 L_3} G_{\bar{\alpha}\bar{\alpha}^\prime J}^{L_1L_2L_3}
    \sum_{\substack{\lambda_b \lambda_c \\
        \lambda_b^\prime \lambda_b^{\prime\prime}}} \sum_{l_1l_2l_3}
    I^{\lambda_b \lambda_b \lambda_b^\prime
      \lambda_b^{\prime\prime}}_{l_1l_2l_3} \times H^{\lambda_b
      \lambda_c \lambda_b^\prime \lambda_b^{\prime\prime}
      L_1L_2L_3}_{\bar{\alpha}\bar{\alpha}^\prime
      l_1l_2l_3}(\Lambda_0) \,.
\end{multline}

The coefficients appearing in the above expression are defined by the
following expression:

\begin{equation}
  S^{JT}_{\bar{\alpha}\bar{\alpha}^\prime} = i^{\bar{l}_{12}+\bar{l} +
    \bar{l}_{12}^\prime +\bar{l}^\prime} (-1)^{\bar{S}_{12} +
    \bar{I}_{12}^\prime - \bar{I} + J + \bar{T}_{12}^\prime }
  \hat{\bar{S}}_{12} \hat{\bar{S}}_{12}^\prime
  \hat{\bar{I}}_{12} \hat{\bar{I}}_{12}^\prime \hat{\bar{I}}
  \hat{\bar{I}}^\prime \hat{\bar{T}}_{12} \hat{\bar{T}}_{12}^\prime
  \hat{T} C_{TM_T,22}^{T^\prime M_T^\prime}
  \NineJ{\bar{T}_{12}}{\frac12}{T}{1}{1}{2}{\bar{T}_{12}^{\prime}}{\frac12}{T^\prime}
  \,,
\end{equation}
\begin{equation}
  G_{\bar{\alpha}\bar{\alpha}^\prime J}^{L_1L_2L_3} = (-1)^{L_1 + L_2
    +L_3} \hat{L}_1^2 \hat{L}_2^2 \hat{L}_3^2
  \SixJ{\bar{I}_{12}^{\prime}}{\bar{I}^{\prime}}{J}{\bar{I}}{\bar{I}_{12}}{L_{1}}
  \NineJ{\bar{S}_{12}^{\prime}}{\bar{S}_{12}}{1}{\bar{l}_{12}^{\prime}}{\bar{l}_{12}}{L_{2}}
  {\bar{I}_{12}^\prime}{\bar{I}_{12}}{L_{1}}\NineJ{\frac{1}{2}}{\frac{1}{2}}{1}{\bar{l}^{\prime}}
  {\bar{l}}{L_{3}}{\bar{I}^\prime}{\bar{I}}{L_{1}}\,,
\end{equation}

\begin{equation}
\begin{split}
  I^{\lambda_b \lambda_c \lambda_b^\prime \lambda_b^{\prime\prime}}_{l_1l_2l_3} =
    & \; (-1)^{\lambda_b + \lambda_c + \lambda_b^\prime +
      \lambda_b^{\prime\prime}} 3^{-\frac{\lambda_b}{2}}
      \widehat{1 - \lambda_b} \; \widehat{1-\lambda_b-\lambda_b^{\prime}}
      \widehat{1-\lambda_c} \widehat{\lambda_b -
      \lambda_b^{\prime\prime}} \\
    & \quad \times  \left[\Bino{2 \lambda_b}{3} \Bino{2\lambda_b^\prime}
      {3 - 2\lambda_b} \Bino{2 \lambda_c}{3}
      \Bino{2\lambda_b^{\prime\prime}}{2\lambda_b + 1}
    \right]^{\frac{1}{2}}
  \hat{l}_1^2 \hat{l}_2 \hat{l}_3 \;
    \CG{1-\lambda_c}{0}{\lambda_b-\lambda_b^{\prime \prime}}{0}{l_2}{0} \;
    \CG{\lambda_c}{0}{\lambda_b^{\prime \prime}}{0}{l_3}{0}
    \NineJ{1-\lambda_c}{\lambda_b - \lambda_b^{\prime\prime}}{l_2}{\lambda_c}
    {\lambda_b^{\prime\prime}}{l_3}{1}{\lambda_b}{l_1}\,,
\end{split}
\end{equation}
\begin{equation}
\begin{split}
  H^{\lambda_b \lambda_c \lambda_b^\prime \lambda_b^{\prime\prime}
    L_1L_2L_3}_{\bar{\alpha}\bar{\alpha}^\prime l_1l_2l_3}(\Lambda_0) =
    & \sum_{\substack{\lambda \lambda^{\prime} \\ \Lambda
      \Lambda^{\prime}}} (-1)^{\lambda^\prime + \Lambda^\prime}
    \hat{\lambda} \hat{\lambda^{\prime}} \hat{\Lambda} \hat{\Lambda}^{\prime}
    \CG{1-\lambda_b -\lambda_b^{\prime}}{0}{\lambda}{0}{\bar{l}_{12}}{0}
    \CG{\lambda_b^{\prime}}{0}{\lambda^{\prime}}{0}{\bar{l}_{12}^\prime}{0}
    \CG{l_2}{0}{\Lambda}{0}{\bar{l}}{0}
    \CG{l_3}{0}{\Lambda^{\prime}}{0}{\bar{l}^\prime}{0} \\
    & \times \sum_{\lambda_3} (-1)^{\lambda_3} \SixJ{1}{1 -
      \lambda_b}{\lambda_b}{\lambda_3}{L_1}{L_2}
      \SixJ{1}{l_1}{\lambda_b}{\lambda_3}{L_1}{L_3}
      \NineJ{1-\lambda_b -
        \lambda_b^\prime}{\lambda}{\bar{l}_{12}}{\lambda_b^\prime}{\lambda^{\prime}}
      {\bar{l}_{12}^\prime}{1-\lambda_b}{\lambda_3}{L_2}
      \NineJ{l_2}{\Lambda}{\bar{l}}{l_3}{\Lambda^{\prime}}{\bar{l}^\prime}{l_1}
      {\lambda_3}{L_3} \\
    & \times \sum_{\lambda_3 \lambda^\prime}
    \widehat{\lambda_3-\lambda_3^{\prime}} \left[C_{2
        \lambda_3^{\prime}}^{2 \lambda_3+1} \right]^{\frac{1}{2}}
    \times C_{\lambda_2 0, \lambda_3-\lambda_3^{\prime} 0}^{\lambda 0}
   C_{\lambda_2 0, \lambda_3^{\prime} 0}^{\lambda^{\prime}0}
      \times
      \SixJ{\lambda_3-\lambda_3^{\prime}}{\lambda_3^{\prime}}{\lambda_3}
      {\lambda^{\prime}}{\lambda}{\lambda_2}
      \\
    & \times \sum_{\lambda_1 \lambda_3^{\prime\prime}}
    \widehat{\lambda_3-\lambda_3^{\prime \prime}}
      \left[C_{2 \lambda_3^{\prime\prime}}^{2
          \lambda_3+1}\right]^{\frac{1}{2}} \times C_{\lambda_1 0,
        \lambda_3-\lambda_3^{\prime \prime} 0}^{\Lambda 0}
      C_{\lambda_1 0, \lambda_3^{\prime \prime} 0}^{\Lambda^{\prime}
        0}
      \times \SixJ{\lambda_3-\lambda_3^{\prime
          \prime}}{\lambda_3^{\prime
          \prime}}{\lambda_3}{\Lambda^{\prime}}{\Lambda}{\lambda_1} \\
    & \times \iiiint \dd k \dd K \dd k^\prime \dd K^\prime
      f_{\lambda_1\lambda_2\lambda_3}\left(k,K,k^{\prime},K^{\prime} \right) \\
    & \qquad \qquad \quad k^{2-\lambda_b-\lambda_b^{\prime}+\lambda_3-\lambda_3^{\prime}}
      K^{2 -\lambda_c + \lambda_b - \lambda_b^{\prime
          \prime}+\lambda_3-\lambda_3^{\prime \prime}}
      {k^\prime}^{1+\lambda_b^{\prime}+ \lambda_3^{\prime}}
      {K^{\prime}}^{1 + \lambda_c + \lambda_b^{\prime
          \prime}+\lambda_3^{\prime \prime}} \\
    & \qquad \qquad \quad P_{\bar{n}_{12}\bar{l}_{12}}(k)P_{\bar{n}\bar{l}}(K)
      P_{\bar{n}_{12}^{\prime}\bar{l}_{12}^{\prime}}(k^{\prime})
      P_{\bar{n}^{\prime}\bar{l}^{\prime}}(K^{\prime})
      f_\text{reg}(k,K;\Lambda_0)
      f_\text{reg}(k^\prime,K^\prime;\Lambda_0) \,,
\end{split}
\end{equation}

\noindent
where we have introduced the shorthand notation
$\hat{x}\equiv\sqrt{2x+1}$.
Here, $C_m^n\equiv n!/[(n-m)!m!]$ denotes the binomial coefficient,
and $P_{nl}(k)$ is the momentum-space HO wave function.
The function arising from the multiple angular expansion is given by

\begin{equation}
\begin{split}
  f_{\lambda_1\lambda_2\lambda_3} \left(k,K,k^{\prime},K^{\prime}
  \right) =
  &\frac{\hat{\lambda}_{1}^{2}\hat{\lambda}_{2}^{2}\hat{\lambda}_{3}^{2}}{8}
  \int_{-1}^{1}\int_{-1}^{1}\int_{-1}^{1}\dd w_{1} \dd w_{2}\dd w_{3}
  P_{\lambda_1}(w_1) P_{\lambda_2}(w_2) P_{\lambda_3}(w_3) \\
  & \times \left(|\vec{k} - \vec{k}^\prime| |\vec{K} - \vec{K}^\prime|
  \right)^{-\lambda_3} \frac{2\times\left(\frac32
    \right)^3}{\vec{q}^2_c \left( \vec{q}_c^2 + m_\pi^2
    \right)^{2}\left(\vec{q}^2_b + m_\pi^2 \right)} \,,
\end{split}
\end{equation}

\noindent
where $P_{\lambda_i}$ is the Legendre polynomial.

\end{widetext}

The next step is to perform the antisymmetrization of the MEs, and
this is conveniently carried out in the Jacobi-HO basis.
The antisymmetrized Jacobi-HO basis MEs are given by

\begin{multline}
  {}_A\la \alpha'; J T^\prime|w_{c_1} | \alpha; J T \ra_A = \\
  \null 6 \sum_{\bar{\alpha}\bar{\alpha}^\prime}
  D_{\alpha^\prime\bar{\alpha}^{\prime}}^{(JT^\prime)}
  D_{\alpha\bar{\alpha}}^{(JT)} \la \bar{\alpha}^\prime; J
  T^\prime|w_{c_1} | \bar{\alpha}; J T \ra \,,
\end{multline}

\noindent
where the matrix $D_{\alpha\bar{\alpha}}^{(JT)}$ is constructed by
diagonalizing the three-body antisymmetrizer

\begin{equation}
  D_{\alpha\bar{\alpha}}^{(JT)} = \sum_{i} \la \alpha; J T|i\ra \la i | \bar{\alpha}; J T \ra \,,
\end{equation}

\noindent
where the index $i$ labels the different antisymmetrized basis states
and does not correspond to a physical quantum number.
Details of construction of the $D$ matrix can be found in
Ref. \cite{Navratil00c}.

The final step is to transform the antisymmetrized matrix elements
from the Jacobi-HO basis to the $JT$-coupled three-particle HO basis
(equivalently, the three-body Slater-determinant basis of HO
single-particle states).
This transformation is performed using the so-called $T$-coefficients,
which involves a sequence of angular-momentum recouplings and
Talmi-Moshinsky transformations.
The explicit construction of the $T$-coefficients can be found in
Ref. \cite{Roth14}.

\bibliographystyle{apsrev}
\bibliography{biblio.bib}

\end{document}